\documentclass[twocolumn,times]{aastex63}

\newcommand{\kms}{{km s$^{-1}$}}

\newcommand{\myemail}{jskang@astro.snu.ac.kr}
\newcommand{\profemail}{mglee@astro.snu.ac.kr}

\usepackage{amsmath}
\hypersetup{linkcolor=red,citecolor=blue,filecolor=cyan,urlcolor=magenta}

\received{November 27, 2020}
\accepted{March 30, 2021}
\submitjournal{ApJ}

\shorttitle{GCSs of Local MCEGs}
\shortauthors{Kang \& Lee}

\begin{document}

\title{
Globular Cluster Systems of Massive Compact Elliptical Galaxies in the Local Universe: 
\\ Evidence for Relic Red Nuggets? 
}

\correspondingauthor{Myung Gyoon Lee}
\email{\myemail,\profemail}

\author[0000-0003-3734-1995]{Jisu Kang} 
\affiliation{Astronomy Program, Department of Physics and Astronomy, 
Seoul National University, 1 Gwanak-ro, Gwanak-gu, Seoul 08826, Republic of Korea}

\author[0000-0003-2713-6744]{Myung Gyoon Lee} 
\affiliation{Astronomy Program, Department of Physics and Astronomy, 
Seoul National University, 1 Gwanak-ro, Gwanak-gu, Seoul 08826, Republic of Korea}

\begin{abstract}
{Nearby massive compact elliptical galaxies (MCEGs) are strong candidates for relic galaxies 
(i.e. local analogs of red nuggets at high redshifts). 
It is expected that the globular cluster (GC) systems of relic galaxies 
are dominated by red (metal-rich) GCs.
NGC 1277 is known as a unique example of such a galaxy in the previous study. 
In this study, we search for GCs in 
12 nearby MCEGs at distances of $\lesssim 100$ Mpc
from the Hubble Space Telescope/Wide Field Camera 3 F814W($I_{814}$)/F160W($H_{160}$) archival images.
We find that most of these MCEGs host a rich population of GCs with a color range of $0.0<(I_{814}-H_{160})_0<1.1$. 
The fractions of their red GCs range from  $f_{RGC} =0.2$ to 0.7 with a mean of $f_{RGC} =0.48\pm0.14$. 
We divide the MCEG sample into two groups: one in clusters and the other in groups/fields. The mean red GC fraction of the cluster MCEGs is $0.60\pm0.06$, which is 0.2 larger than the value of the group/field MCEGs, $0.40\pm0.10$.
The value for the cluster MCEGs is $\sim$0.3 larger than the mean value 
of giant early-type galaxies with similar stellar mass in the Virgo cluster ($f_{RGC} =0.33\pm0.13$).   
Our results show that most of the MCEGs in our sample are indeed relic galaxies. 
This further implies that a majority of the red GCs in MCEGs are formed early in massive galaxies 
and that most MCEGs in the local universe have rarely undergone mergers after they became red nuggets about 10 Gyr ago. 
}
\end{abstract}

\section{Introduction}

According to the current understanding of galaxy evolution, 
massive early-type galaxies (ETGs) form in two-phases 
\citep[e.g.][and references therein]{kho06,van10,ose10,ose12,van14,wel16,nab17,tof14,tof17}. 
They grow in size and mass from massive compact red progenitors through successive mergers. 
These progenitors are 
similar to `red nugget' galaxies found at high redshift at $z=2-3$. 
Red nugget galaxies are massive, compact, and composed of old stellar populations 
showing little evidence of recent star formation. 

In the first phase of formation and evolution,
red nugget galaxies might have underwent the following steps:
(1) started with accretion-driven violent disc instability, leading to fast rotation; 
(2) contracted via dissipative process, forming compact, star-forming blue nuggets; and
(3) star formation in the blue nuggets quenched by feedback due to AGN activity. 
The blue nuggets became red nuggets about 10 Gyr ago \citep[e.g.][]{dek14,alm17,tor18}.

In the second phase of formation and evolution,  
most red nugget galaxies grow in size and mass by a number of minor mergers.
They become larger ETGs with higher mass, as often seen in the local universe.
According to deep observations of high redshift galaxies and simulations on size evolution, 
massive compact ETGs at $z=2-3$ have effective radii of $\approx$ 
1 kpc 
and they grow by a factor of 5 to 6 in size until present day \citep[e.g.][]{kho06,ose12,van14,fur17,lap18}.

%

However, if some of the red nuggets at high redshift rarely experience mergers during their evolution, 
they would appear as relic red nuggets in the local universe.
Indeed, 
a small number of the relic red nugget candidates are found recently in the local universe 
\citep[e.g.][and references therein]{yil17}. 
These galaxies are good candidates for relic galaxies. 
\citet{tru14} defined a relic galaxy as 
``an object that was formed in the early phases of the universe (i.e., $z > 2$) 
and has remained unaltered
(i.e., without significant gas or stellar accretion) since its initial formation'', 
and pointed out that these relic galaxies would appear today as 
``old massive galaxies which have the same structural properties 
as the population of massive galaxies at high-$z$'' (i.e. old massive compact galaxies). 
\citet{tru14} suggested selection criteria for relic galaxy candidates 
in the local universe: 
old age ($>10$ Gyr), high stellar mass ($M_* \ga 10^{11} M_\sun$), 
and compact size ($R_e \la 1.5$ kpc).


There are several confirmed relic galaxies in the local universe. First, 
NGC 1277, a peculiar S0 galaxy at a distance of 73 Mpc in the Perseus cluster, 
has been 
confirmed as 
a relic galaxy \citep{tru14}.  
From the analysis of imaging and spectral data of the galaxy light, 
\citet{tru14} found that NGC 1277 is old ($>10$ Gyr), 
massive ($M_*=1.2\pm0.4 \times 10^{11} M_\sun$), and compact ($R_e = 1.2$ kpc).
They also estimated the metallicity of the central region in NGC 1277 to be above solar 
([Fe/H]$=0.20\pm0.04$ and [$\alpha$/Fe]$=0.4\pm0.1$).
NGC 1277 is rotating very 
quickly with a rotation velocity of $v_{rot}\approx300$ \kms, 
and it also shows a high central velocity dispersion of $\sigma_v=355$ \kms \citep{yil17}. 
Later, two more relic galaxies at distances of $\lesssim 100$ Mpc, MRK 1216 and PGC 32873, 
were confirmed by \citet{fer17}. 
They pointed out that these two galaxies host {\"u}bermassive black holes ({\"U}MBH). 
They also inferred that these two galaxies have steep stellar initial mass function (IMF) profiles.
Most recently, \citet{spi20} confirmed two more relic galaxies at intermediate redshift, 
KiDS J0224-3143 ($z=0.38$) and KiDS J0847+0112 ($z=0.18$).


\citet{yil17} presented a study of 16 compact elliptical galaxies 
thought to be good candidates for relic galaxies.
These compact elliptical galaxies were selected 
from the Hobby-Eberly Telescope Massive Galaxy Survey \citep[HETMGS;][]{van15}. 
This survey aims to find nearby galaxies 
that are suitable for a dynamical black hole mass measurement with high spatial resolution imaging follow-up. 
From the survey, 16 compact elliptical galaxies were selected 
with tight criteria in sphere-of-influence $\ge0\farcs05$ and $R_e\le2$ kpc.
The structural and dynamical properties of these selected galaxies were studied 
based on Hubble Space Telescope (HST) F160W images and wide-field integral field unit data.
They are strong candidates for relic 
red nuggets
in the sense that they follow the stellar mass--size relation 
for massive compact red galaxies at $z\approx2$, 
while they show a significant deviation from the 
relation 
for massive ETGs in the local universe \citep{yil17}. 
Most 
have old ages ($\ga 10$ Gyr) and high metallicities above solar.
NGC 1277
, MRK 1216 and PGC 32873 are 
also included in the list of \citet{yil17}.


Although numerous studies have 
provided useful information 
about structural, photometric, and kinematic properties of the massive compact elliptical galaxies (MCEGs) 
from the analysis of their integrated galaxy light \citep[e.g.][and references therein]{yil15,yil16, yil17},    
much less is known about globular cluster (GC) systems in these galaxies. 
GCs are excellent tools to trace how their host galaxies evolve. 
In general, massive ETGs often host two subpopulations of GCs: 
blue (metal-poor) GCs and red (metal-rich) GCs 
\citep[see][and references therein]{bro06,har17,for18}. 
Red GCs are believed to have been formed in the progenitors of massive galaxies, 
while 
blue GCs are believed to have been formed in low mass galaxies 
and accreted to the massive galaxies through mergers 
\citep[][and references therein]{bro06,lee10a,har17,for18}. 

The GCs in the Milky Way (MW) Galaxy are the best targets to estimate the ages of the GCs. The ages of the GCs in the MW Galaxy 
based on the recent isochrone fitting of the main sequences \citep[e.g.][]{oli20} 
range from 11.7 to 13.5 Gyrs, with a mean value of $12.3\pm0.4$ Gyr.
The mean age of the metal-rich (red) GCs ([Fe/H] $>-0.85$), $12.12\pm0.32$ Gyr, 
is slightly smaller than that of the metal-poor (blue) GCs ([Fe/H] $<-0.85$), $12.86\pm0.36$ Gyr.
The age scatter in each subpopulation, as well as the age difference between the two subpopulations, 
is smaller than 1 Gyr, showing that they are formed almost simultaneously.
Thus, the GCs are an efficient tracer of relic galaxies. 

It is expected that GC systems in relic galaxies are dominated by red GCs 
if the relic galaxies have rarely undergone mergers 
after they became massive compact red galaxies at about 10 Gyr ago. 
Recently, from HST F475W($g_{475}$)/F850LP($z_{850}$) photometry, 
\citet{bea18} found that the GC system in NGC 1277 is dominated by red GCs 
(showing a red GC fraction of 
$f_{RGC}>83\%$), 
and they estimated that the total stellar mass accretion in this galaxy is minor ($< 10$\%).
From these results, they suggested that NGC 1277 is indeed a genuine relic galaxy.
To date, NGC 1277 is the only MCEG whose GC system 
has been studied to estimate the red GC fraction.

In this study, we investigate GC systems in a large sample of nearby MCEGs. 
We expect to draw a conclusion 
whether the GC system of NGC 1277 is special 
or if most MCEGs follow NGC 1277 in the fraction of red GCs. 
The primary goals of this study are (1) to select GCs in each galaxy,
(2) to estimate the fraction of red GCs in each galaxy,
and (3) to investigate any differences in the GC systems 
(especially the fraction of the red GCs) 
between the MCEGs and the massive ETGs in the local universe.

\begin{deluxetable*}{lrccclccl}
\tabletypesize{\footnotesize}
\tablecaption{Basic Properties of the Target MCEGs and Reference Galaxies \label{tab_list} }
\tablewidth{0pt}
\tablehead{
\colhead{Galaxy} & \colhead{$D$ [Mpc]} &  \colhead{$R_{e,circ}$ [kpc]} & \colhead{$b/a$} & \colhead{$\sigma_c$ [\kms]} & \colhead{log($M_*/M_\sun$)} & 
\colhead{$B$ [mag]} &
\colhead{$A_I$ and $A_H$} & \colhead{$f_{RGC}$} \\ 
\colhead{(1)} & \colhead{(2)} & \colhead{(3)} & \colhead{(4)} & \colhead{(5)} & \colhead{(6)} & \colhead{(7)} & \colhead{(8)} & \colhead{(9)} 
}
\startdata
UGC 3816 & $51\pm1$ & $1.8\pm0.1$ & 0.69 & $251\pm7$ & 10.96$^{+0.06}_{-0.04}$ & 
13.46 & 0.095, 0.032 & $0.51^{+0.12}_{-0.10}$ \\
NGC 0384 & $59\pm1$ & $1.5\pm0.1$ & 0.68 & $240\pm5$ & 10.96$^{+0.05}_{-0.05}$ & 
13.67 & 0.096, 0.032 & $0.45^{+0.08}_{-0.09}$ \\
NGC 1281 & $60\pm1$ & $1.3\pm0.1$ & 0.64 & $263\pm6$ & 11.00$^{+0.08}_{-0.08}$ & 
13.74 & 0.256, 0.085 & $0.70^{+0.09}_{-0.09}$ \\
NGC 1270 & $69\pm1$ & $1.9\pm0.1$ & 0.68 & $376\pm9$ & 11.31$^{+0.10}_{-0.12}$ & 
13.43 & 0.253, 0.084 & $0.61^{+0.06}_{-0.02}$ \\
PGC 70520 & $72\pm1$ & $1.2\pm0.1$ & 0.49 & $259\pm8$ & 10.95$^{+0.10}_{-0.12}$ & 
14.17 & 0.147, 0.049 & $0.30^{+0.07}_{-0.07}$ \\
NGC 0472 & $74\pm1$ & $2.0\pm0.1$ & 0.72 & $252\pm7$ & 11.07$^{+0.06}_{-0.11}$ & 
13.73 & 0.075, 0.025 & $0.41^{+0.12}_{-0.07}$ \\
NGC 2767 & $74\pm1$ & $1.9\pm0.1$ & 0.75 & $247\pm9$ & 11.12$^{+0.09}_{-0.08}$ & 
14.36 & 0.029, 0.010 & $0.37^{+0.01}_{-0.05}$ \\
NGC 1271 & $80\pm2$ & $1.4\pm0.1$ & 0.43 & $302\pm8$ & 11.06$^{+0.07}_{-0.07}$ & 
14.30 & 0.254, 0.085 & $0.60^{+0.05}_{-0.05}$ \\
UGC 2698 & $89\pm2$ & $3.1\pm0.1$ & 0.73 & $351\pm8$ & 11.58$^{+0.01}_{-0.03}$ & 
13.61 & 0.224, 0.075 & $0.56^{+0.04}_{-0.04}$ \\ 
MRK 1216 & $94\pm2$ & $2.3\pm0.1$ & 0.58 & $335\pm6$ & 11.34$^{+0.11}_{-0.10}$ & 
14.44 & 0.050, 0.017 & $0.52^{+0.09}_{-0.04}$ \\ 
PGC 11179 & $94\pm2$ & $1.8\pm0.1$ & 0.66 & $292\pm7$ & 11.16$^{+0.06}_{-0.08}$ & 
14.51 & 0.287, 0.096 & $0.56^{+0.03}_{-0.11}$ \\
PGC 32873 & $112\pm2$ & $1.9\pm0.1$ & 0.53 & $308\pm9$ & 11.28$^{+0.04}_{-0.04}$ & 
14.98 & 0.019, 0.006 & $0.24^{+0.06}_{-0.00}$ \\ 
\hline
NGC 1399 & $20\quad\;\;\,$ & $11.1\qquad\;\;\;\:$ & 0.91 & $332\pm5$ & $11.41$ & 10.60 & 0.019, 0.006 & $0.72^{+0.06}_{-0.10}$ 
\\ 
NGC 4874 & $100\quad\;\;\,$ & $22.7\qquad\;\;\;\:$ & 0.87 & $272\pm4$ & $11.76$ & 12.63 & 0.014, 0.005 & $0.55^{+0.08}_{-0.08}$ 
\\ 
\enddata
\tablecomments{(1) Galaxy name; (2) Distances; (3) Effective radii within circularized aperture; 
(4) The ratio of minor axis and major axis; (5) Central velocity dispersion; (6) Stellar mass 
derived from simple stellar population models for Salpeter-like stellar IMF (with a slope of $\Gamma=2.35$); (7) B-band magnitudes; (8) Foreground extinction for F814W($I_{814}$) and F160W($H_{160}$) bands; (9) Fractions of red GCs derived in this study ($R<10R_{e,circ}$ for the target MCEGs, $R\lesssim1R_e$ for NGC 1399 and $R\lesssim2R_e$ for NGC 4874)
}
\tablerefs{(1--6) \citet{yil17} for the target MCEGs; (1,2,3,6) ACSFCS for NGC 1399, ACSCCS for NGC 4874; 
(4,5) HyperLeda for NGC 1399 and NGC 4874; (7) HyperLeda; (8) \citet{sch11}; (9) This study}
\end{deluxetable*}


We select target MCEGs 
from the list of 16 compact elliptical galaxies in \citet{yil17}. 
It is an excellent resource 
because HST F814W($I_{814}$)/F160W($H_{160}$) images are available for most of them.
From the list, we exclude NGC 3990, NGC 1282 and PGC 12562 
because they have a stellar mass much lower than the other MCEGs.
There are no F160W images available for NGC 1277 so we exclude it as well.
Thus, we select 
12 MCEGs as listed in Table \ref{tab_list}. 
Basic properties of the target MCEGs given by \citet{yil17} are summarized in the table.
The ranges of their physical parameters are: 
circularized effective radii of $R_{e,circ}=$ 
1.2 to 3.1 kpc, 
central velocity dispersion of $\sigma_c=$ 
240 to 376 \kms, 
and stellar mass of log$(M_*/M_\sun) =$ 
10.95 to 11.58. 

After submitting this paper, \citet{ala21} presented a similar study of the MCEG GCs. They used the same MCEG targets as this 
work to study the general properties of the GCs in these galaxies, but they did not mention any subpopulations of the GCs. 
They did not provide any tables for the photometric data we can use for quantitative comparison, so we compared visually the 
color-magnitude diagrams (CMDs) and radial number density profiles of the GCs with \citet{ala21}, finding that both are consistent with each other. 

This paper is organized as follows.
Section \ref{data} describes the basic properties of the data, how we reduce the data, 
how we select GC candidates in each galaxy, 
and how we derive the total magnitudes and colors of the GC candidates. 
In Section \ref{result} we present the 
CMDs 
and color distributions of the GCs in each galaxy, 
and how we estimate the fraction of red GCs in each galaxy. 
Then we show the spatial and radial distributions of the GCs in each galaxy.
Section \ref{discuss} discusses how the fractions of red GCs vary 
depending on sizes, magnitudes, and stellar masses
of their host galaxies, 
and its implications on the origin of GCs and the evolution of MCEGs. 
Our results and conclusion are summarized in the final section.

\begin{deluxetable}{lrrrc}
\tabletypesize{\footnotesize}
\tablecaption{HST Data for the Target MCEGs and Reference Galaxies \label{tab_data} }
\tablewidth{0pt}
\tablehead{
\colhead{Galaxy} & \colhead{T(F475W)} & \colhead{T(F814W)} &\colhead{T(F160W)} & \colhead{PID} 
\\ \colhead{} & \colhead{[s]} & \colhead{[s]} & \colhead{[s]} & \colhead{}
}
\startdata
UGC 3816  & - & 696 & 1500 & 13050  \\
NGC 0384  & - & 482 & 1350 & 13050  \\ 
NGC 1281  & - & 495 & 1350 & 13050  \\
NGC 1270  & - & 495 & 1350 & 13050  \\
PGC 70520 & - & 489 & 1350 & 13050  \\
NGC 0472  & - & 482 & 1350 & 13050  \\
NGC 2767  & - & 635 & 1500 & 13050  \\
NGC 1271  & - & 495 & 1350 & 13050  \\
UGC 2698  & - & 805 & 1350 & 13416  \\
MRK 1216  & - & 500 & 1350 & 13050  \\
PGC 11179 & - & 495 & 1350 & 13050  \\
PGC 32873 & - & 538 & 1500 & 13050  \\
\hline
NGC 1399 & 680 & 1224 & 1197 & 10911, 11712 \\ 
NGC 4874 & 5071 & 10425 & 10791 & 10861, 11711 \\
\enddata
\end{deluxetable}

\section{Data Reduction and GC Candidate Selection} \label{data}

\subsection{Data}

We use Wide Field Camera 3 (WFC3)/UVIS F814W($I_{814}$) and WFC3/IR F160W($H_{160}$) images 
of the target galaxies 
from the HST archive (PI: van den Bosch, PID: 13050/13416).
Basic information of the images are listed in Table \ref{tab_data}.
For each galaxy, 
the total exposure times are $\approx$ 500 s for F814W and $\approx$ 1400 s for F160W.
F814W images cover a field of view of $2\farcm7 \times 2\farcm7$, 
which is larger than the field of view of F160W images, $2\farcm27 \times 2\farcm05$. 
We use only F160W fields which are spatially overlapped with F814W fields for the following analysis. 

For reference, we 
analyze 
Advanced Camera for Surveys (ACS)/Wide Field Channel (WFC) F475W($g_{475}$), F814W($I_{814}$) and WFC3/IR F160W($H_{160}$) archival images 
of two local massive ETGs, NGC 1399 and NGC 4874.
More details of 
these images are explained in \citet{bla12} and \citet{cho16}.

\begin{figure*}
\centering
\includegraphics[scale=0.8]{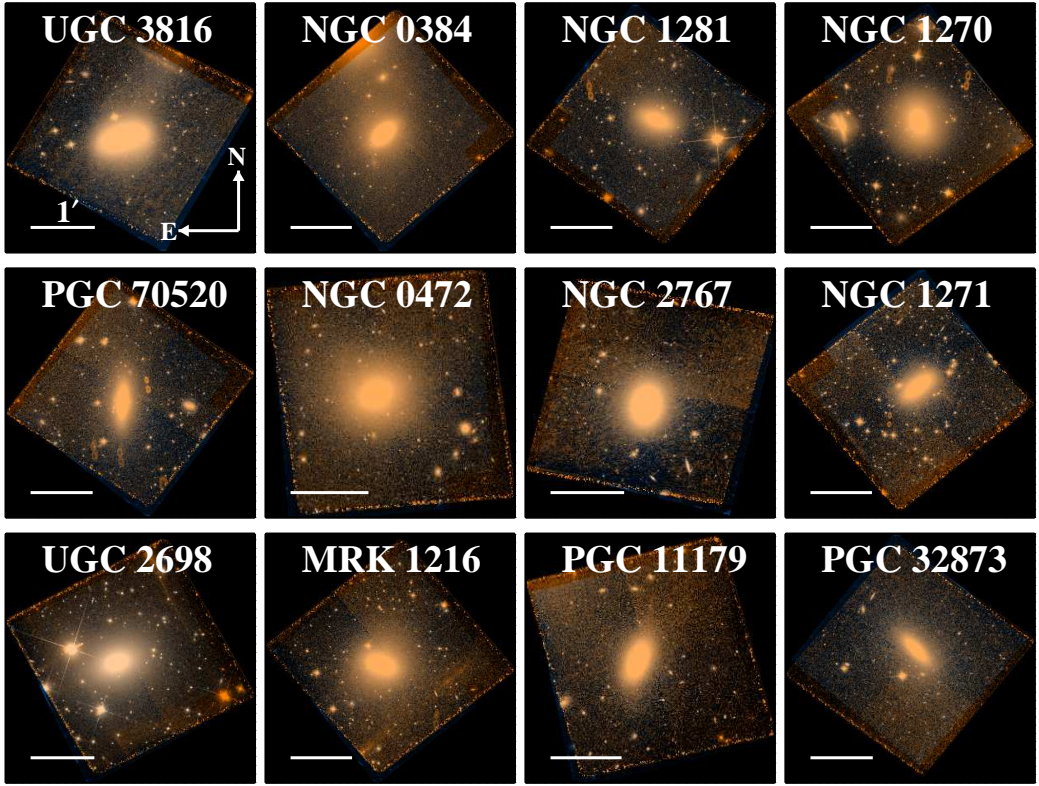}
\caption{
Pseudo-color stamp maps of the 
12 target MCEGs.
WFC3/UVIS F814W images and WFC3/IR F160W images are combined.
North is up and east to the left. 
One arcminute scale bars are marked on the bottom left side.
}
\label{fig_maps}
\end{figure*}

\subsection{Data Reduction}

We reduce the HST images following the procedures we applied in our previous studies 
of GCs in nearby galaxies \citep[e.g.][]{lee16a,lee19}.
We download individual frame images of each galaxy
(\texttt{\_flt} for WFC3/IR and \texttt{\_flc} for the rest) 
from the MAST archive\footnote{\url{http://archive.stsci.edu/hst/}}.
They are aligned and combined within the same filter using \texttt{DrizzlePac} \citep{gon12}. 
The image scale of the combined image is $0\farcs04$ pixel$^{-1}$ for WFC3/UVIS F814W, 
$0\farcs05$ pixel$^{-1}$ for ACS/WFC F475W and F814W, 
and $0\farcs10$ pixel$^{-1}$ for WFC3/IR 
F160W. 
We adopt \texttt{final\_pixfrac}=0.8 and \texttt{final\_pixscale}=0.10 for WFC3/IR F160W 
following the values used for drizzling NGC 1399 and NGC 4874 images \citep{bla12,cho16}.
FWHM values of the point sources are around 2 pixels for every image. 
Figure \ref{fig_maps} displays 
pseudo-color maps of our 
12 target MCEGs. 
Similar maps of the F160W images of 16 compact elliptical galaxies including our target galaxies 
are given in Figure 1 of \citet{yil17}.

For better detection of GC candidates from the images, 
we detect sources after subtracting the contribution of the diffuse galaxy light 
from the original images.
We produce a smooth, filtered image from the drizzled image of each galaxy 
using 
a ring median filter with \texttt{IRAF/RMEDIAN} \citep{sec95}.
Then we subtract the smooth image from the drizzled image 
to produce a galaxy-light-subtracted image,
and use the resulting image for source detection.
We use the ring median filters 
with inner radius of 16 pixels and outer radius of 18 pixels.
If the radius of the ring median filter is too small, 
then the flux of the GC candidates could be underestimated.
If the radius is too large, 
then the galaxy light is not fully filtered.
Considering these, we set the inner filter radius as 8 times 
the FWHM value.

We run \texttt{Source Extractor} to detect the sources in the galaxy-light-subtracted images
with 2 $\sigma$ threshold \citep{ber96}.
We set \texttt{BACK\_SIZE} = 32 pixels and \texttt{BACKPHOTO\_TYPE} = LOCAL 
for better detection of GC candidates near bright sources.
Other settings are not changed from the default.
To remove the spurious sources, 
we detect sources separately for the F814W and F160W images 
and only select sources that appear in both images with a matching tolerance of 0\farcs2.
For the selected sources, 
we obtain auto magnitudes 
and aperture magnitudes 
with various aperture radii ranging from 3 to 5 pixels. 
To get total magnitudes and colors, we correct the small aperture magnitudes 
by applying aperture correction as described in Section \ref{apcor}.

We adopt AB magnitudes to be consistent with previous studies of the reference galaxies 
NGC 1399 and NGC 4874 \citep{bla12, cho16}. 
Calibration of the instrumental magnitudes are done using the zeropoints 
from the online WFC3 table\footnote{\url{https://www.stsci.edu/hst/instrumentation/wfc3/data-analysis/photometric-calibration/}} or 
from the online ACS Zeropoint Calculator\footnote{\url{https://acszeropoints.stsci.edu/}}.
For the case of MCEGs, WFC3/UVIS F814W zeropoints for an infinite aperture is 
25.125 (average of UVIS1 25.139 and UVIS2 25.110).
ACS F475W/F814W zeropoints for an infinite aperture are 26.055 and 25.939 for NGC 1399, 
and 26.057 (average of 26.054 and 26.060) and 25.949 for NGC 4874.
WFC3/IR F160W zeropoint for an infinite aperture is 25.946 for all cases.

\subsection{Selection of GC Candidates} \label{gcselect}

\begin{figure*}
\centering
\includegraphics[scale=0.8]{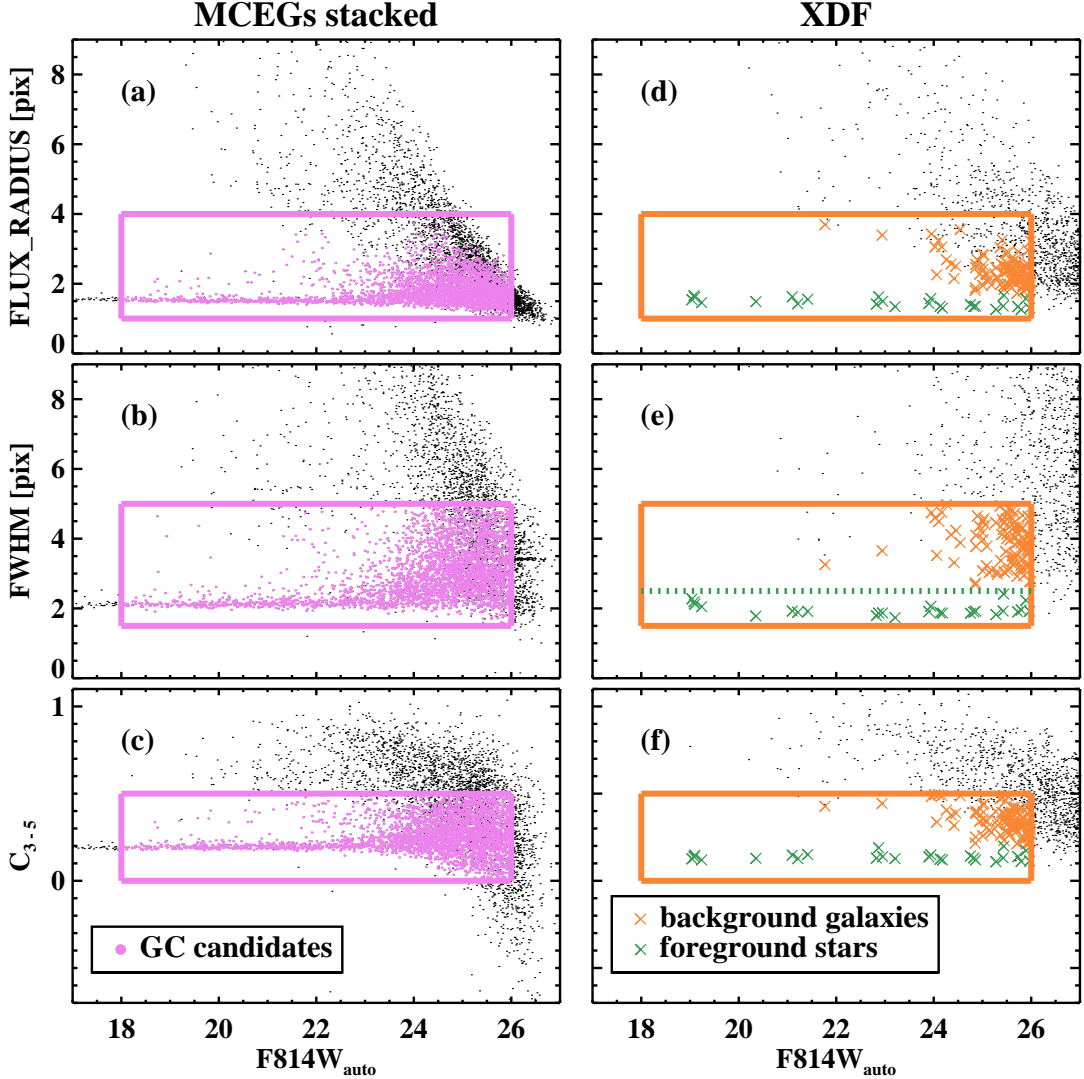}
\caption{
(Left panels) Effective radius (\texttt{FLUX\_RADIUS}), FWHM, and concentration parameter ($C_{3-5}$) 
versus F814W auto magnitude of the sources detected in the images of the target MCEGs. 
We plot this diagram after stacking all the results from each MCEG. 
We mark the selection criteria for GC candidate in pink boxes and selected GC candidates in pink dots.
We select the candidates satisfying all three size-related conditions at the same time.
(Right panels) Same plots as left panels but for the sources detected in XDF images. 
We apply the same selection criteria as shown in orange boxes in order to calculate the background contamination.
Orange crosses denote 
background sources and green crosses denote foreground stars with FWHM $<2.5$.
}
\label{fig_gcselect}
\end{figure*}

It is expected that GCs in the target galaxies appear 
as point-like sources or slightly resolved sources 
in the F814W/F160W images \citep{har09,lee16a,lee19}. 
Therefore, we first select compact sources from the catalog of the detected sources 
using two \texttt{Source Extractor} output parameters, 
effective radius (\texttt{FLUX\_RADIUS}) and FWHM (\texttt{FWHM\_IMAGE})
as well as magnitude concentration parameter ($C$).
The magnitude concentration parameter used in this study is derived 
from the difference between two different aperture magnitudes 
with aperture radii of 3 pixel and 5 pixel in F814W images: $C_{3-5}=F814W(r=3 {\rm pix}) - F814W(r=5 {\rm pix})$.
To set the parameter range for selecting compact sources, 
we stack the photometric results of our target MCEGs as shown in Figure \ref{fig_gcselect}.

In the left panel of Figure \ref{fig_gcselect}, 
we display the values of effective radius, FWHM, and concentration parameter 
versus F814W auto magnitude of the sources detected in our 
12 MCEG images. 
Note that the bright narrow sequences at \texttt{FLUX\_RADIUS} $\approx1.5$ pixels, \texttt{FWHM\_IMAGE} $\approx2.0$ pixels, and $C_{3-5}\approx0.2$
appear in 
each plot, 
showing that they are point sources. These point sources are mostly foreground stars. 
Moreover, these sources show color-magnitude relations similar to the foreground stars as well 
(see Figures \ref{fig_cmd01} to \ref{fig_cmd13}). 
However, these sequences become unclear as the magnitudes 
become fainter, 
especially in the magnitude range where GC candidates are mostly located.
Therefore, we select GC candidates including these narrow sequences and then subtract the foreground contamination later 
as described in Section \ref{frgc}.
We adopt the conditions for GC candidate selection in F814W images: 
$18.0<$ \texttt{MAG\_AUTO} $<26.0$, 
$1.0<$ \texttt{FLUX\_RADIUS} $<4.0$, 
$1.5<$ \texttt{FWHM\_IMAGE} $<5.0$, and 
$0.0<C_{3-5}<0.5$.
The bright limit is set considering the magnitude of the brightest GCs 
and the faint limit is set considering the magnitude limit. 
The magnitude range is initially broadly set and will be narrowed later. 
The ranges for the other three size parameters are very similar to those from \citet{bla12,cho16,lee18}.
We also consider flags 
to remove any saturated sources or sources near the image boundary 
and ellipticity 
to remove any elongated sources: 
\texttt{FLAGS} $<4$, \texttt{ELLIPTICITY} $<0.56$ for both filters.
We decide to use the cut at 0.56 where a discontinuity in the ellipticity distribution of compact sources is seen. 

\begin{figure*}
\centering
\includegraphics[scale=0.8]{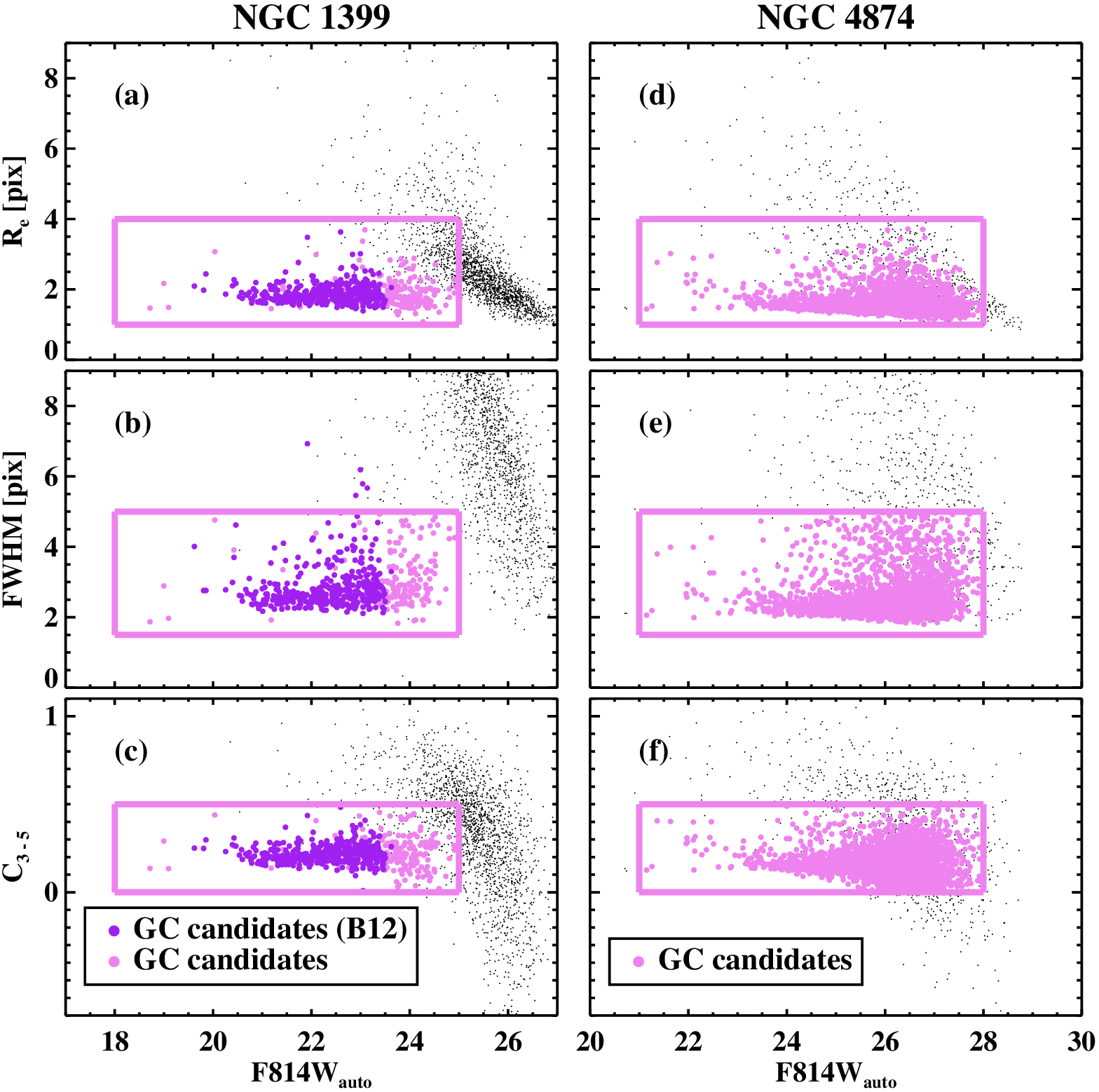}
\caption{
Same plots as Figure \ref{fig_gcselect} 
but for the results of NGC 1399 (left panels) and NGC 4874 (right panels). 
We mark the GC candidates of each galaxy in pink dots 
which were selected with the same selection criteria as target MCEGs.
GC candidates of NGC 1399 provided by \citet{bla12} are marked in purple dots. 
}
\label{fig_gcselect2}
\end{figure*}

To estimate the contamination by background sources, 
we display the results of the sources detected in the Hubble eXtreme Deep Field (XDF) images
in the right panel of Figure \ref{fig_gcselect}. 
We use ACS/WFC F814W data with an image scale of $0\farcs03$ pixel$^{-1}$ and WFC3/IR F160W data with an image scale of $0\farcs06$ pixel$^{-1}$ provided by XDF Data Release 1.0 \citep{ill13}.
Considering the image scale difference between our MCEG data and the XDF data, 
we convert \texttt{FLUX\_RADIUS}, \texttt{FWHM\_IMAGE}, and $C_{3-5}$ obtained from the XDF data 
to have the same angular scales as those of our MCEG data. 
Then we apply the same selection criteria for the GC candidates to calculate and subtract contamination caused by background galaxies.
Most XDF sources selected with the same selection criteria for the GC candidates 
are foreground or background objects. 
To select only the background objects, we exclude the foreground objects with FWHM $< 2.5$ pixels.

For comparison, we also select GC candidates of NGC 1399 and NGC 4874 
with the same selection criteria as shown in Figure \ref{fig_gcselect2}.
F814W auto magnitude range for selecting GC candidates is 
$18.0<$ \texttt{MAG\_AUTO} $<25.0$ for NGC 1399 and $21.0<$ \texttt{MAG\_AUTO} $<28.0$ for NGC 4874.
At the distance to NGC 1399 \citep[$(m-M)_0=31.5$;][]{bla09} and NGC 4874 \citep[$(m-M)_0=35.0$;][]{car08}, 
the magnitude range for the GCs ($M_I > -13$ mag) is expected as F814W $>$ 18.5 mag for NGC 1399 and 22 mag for NGC 4874. 
Considering this and photometric limiting magnitudes of the data, 
we set the magnitude ranges to include all GC candidates with different values.
\citet{bla12} provided a catalog of their photometry for NGC 1399 GCs, so we overlay the results 
and confirmed that most of the GC candidates in \citet{bla12} are within our criteria.
\citet{cho16} did not provide a photometric catalog for NGC 4874 GCs 
so we could not directly compare the results.
Instead, an indirect comparison from the color-color diagram is shown in Section \ref{n4874}.

\subsection{Total Magnitude and Color of the GC Candidates} \label{apcor}

To derive the total magnitude and color of the GC candidates, 
we conduct the following procedures.
First, we derive the F160W total magnitude ($H_{160,tot}$) 
rather than the F814W total magnitude 
because the signal to noise ratio of the F160W images 
is 
larger than that of the F814W images.
Using the value of the F160W zeropoint 
for an aperture with radius $r=0\farcs4$ (4 pixels), 25.755, 
$H_{160,tot}$ can be derived as follows: $H_{160,tot}=F160W(r={\rm 4 pix})-25.946+25.755$.

Second, we derive $(I_{814}-H_{160})_{tot}$ color using the small aperture magnitudes 
to obtain 
a larger signal to noise ratio.
We select the small aperture sizes 
considering the Point Spread Function (PSF) encircled energy fraction (EEF)
from the online WFC3 table mentioned above.
The PSF encircled energy fraction of the WFC3/IR F160W  
within the aperture radius $r=0\farcs3$ (3 pixels) is EEF$=$0.789
which is very similar 
to that of 
WFC3/UVIS F814W 
within the aperture radius $r=0\farcs16$ (4 pixels), EEF$=$0.795. 
Therefore, we derive $(I_{814}-H_{160})_{tot}$ color as follows:
$(I_{814}-H_{160})_{tot}=F814W(r={\rm 4 pix})-F160W(r={\rm 3 pix})+2.5{\rm log}(0.795/0.789)$.
We assume that 
the color gradient outside the small aperture region of the GC candidates is negligible 
and that the encircled energy distribution of the GC candidates 
is similar to that of the PSF.

Finally, we derive the F814W total magnitude ($I_{814,tot}$) by combining the previous results: 
$I_{814,tot}=H_{160,tot}+(I_{814}-H_{160})_{tot}$. 
We correct the observed magnitudes and colors 
for the foreground extinction using the information in \citet{sch11}, 
as listed in Table \ref{tab_list}.
We use `0' subscripts for the foreground extinction-corrected magnitudes 
and colors: 
$I_{814,0}$, $H_{160,0}$, and $(I_{814}-H_{160})_0$.
We consider the internal extinction inside each of the target galaxies to be negligible, 
if any, because they are ETGs.

We follow the same procedures as we used for the MCEGs 
to derive the total magnitude and color of the GC candidates in NGC 1399 and NGC 4874. 
The PSF encircled energy fraction of the ACS F475W and F814W 
within the aperture radius $r=0\farcs15$ (3 pixels) is EEF$=$0.794 and 0.770,  respectively\footnote{\url{https://www.stsci.edu/hst/instrumentation/acs/data-analysis/aperture-corrections}}.
Therefore, we derive the colors as follows:
$(I_{814}-H_{160})_{tot}=F814W(r={\rm 3 pix})-F160W(r={\rm 3 pix})+2.5{\rm log}(0.770/0.789)$ and
$(g_{475}-I_{814})_{tot}=F475W(r={\rm 3 pix})-F814W(r={\rm 3 pix})+2.5{\rm log}(0.794/0.770)$.
Foreground extinction corrections are also applied.

To estimate the completeness of the GC candidate detection, we perform artificial star tests using \texttt{IRAF/ARTDATA}. For each galaxy, we generate 100 images with 1000 artificial point sources with FWHM$=2.0$ pixels and $(I_{814}-H_{160})=0.3$, a mean color of blue GCs (see Figures \ref{fig_cmd01} to \ref{fig_cmd13}). Similarly, we add artificial sources with $(I_{814}-H_{160})=0.7$, a mean color of red GCs to see any completeness difference between the two populations. At $I_{814}=25.0$ mag, the completeness is about 80\% for both blue and red GC cases. The difference between the blue and red GC cases is smaller than 1\% and the error for the red GC fraction due to incompleteness is estimated to be smaller than 1\%. 


\section{Results} \label{result}

\begin{figure*}
\centering
\includegraphics[scale=0.8]{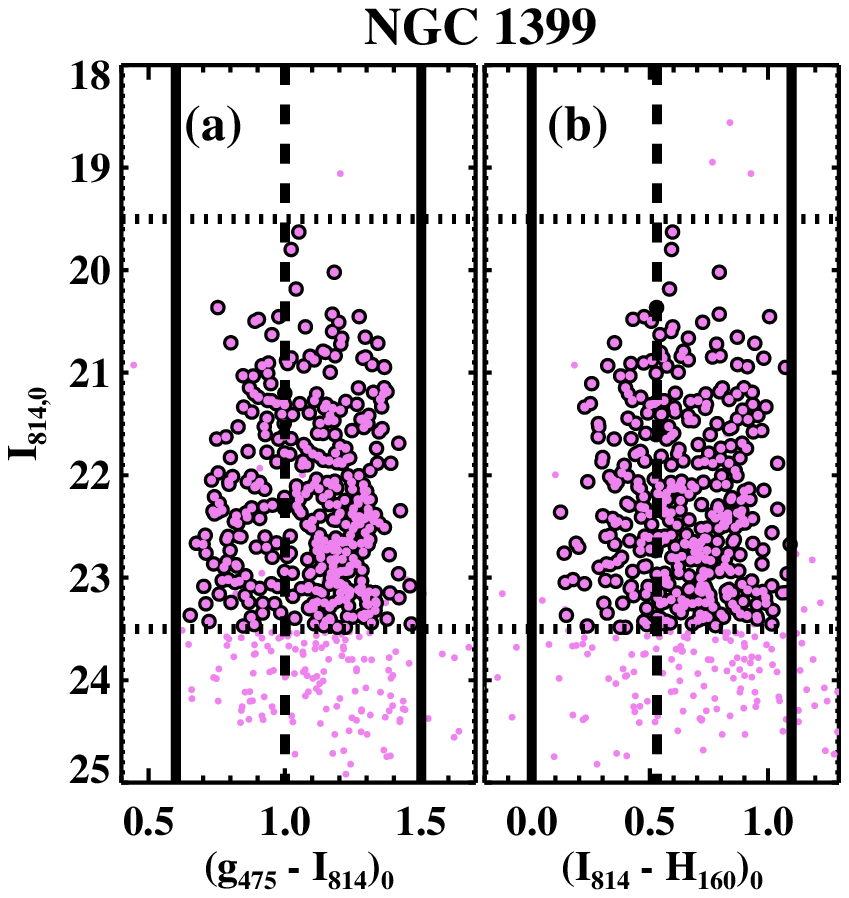}
\includegraphics[scale=0.8]{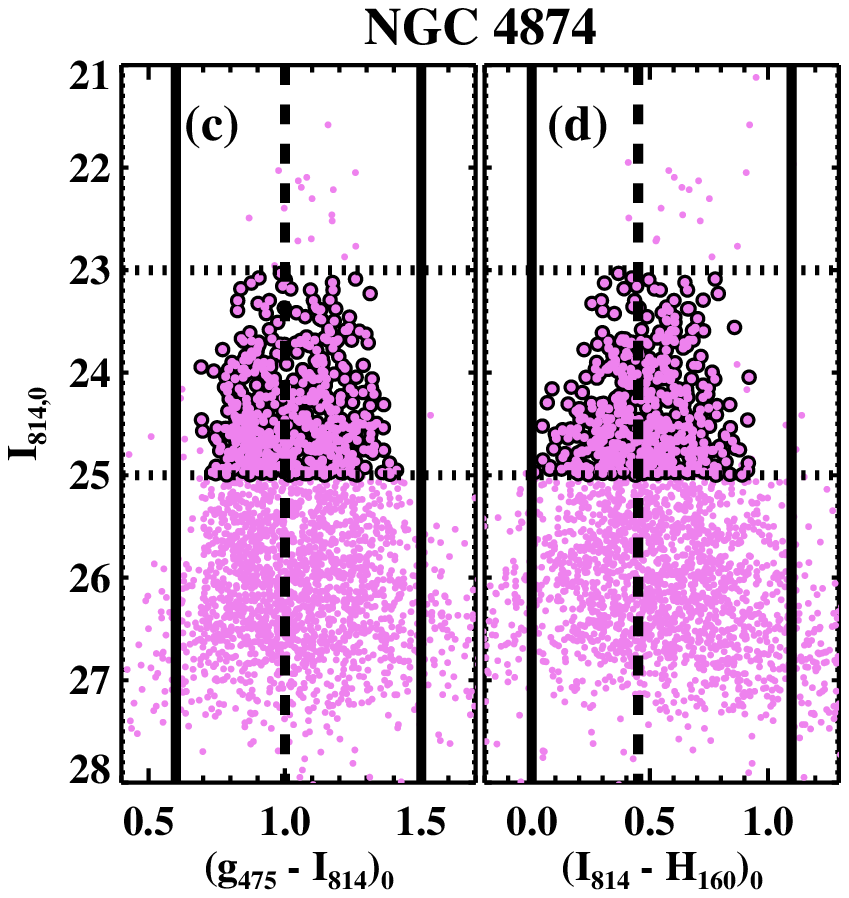}
\caption{
CMDs of the GC candidates in (a,b) NGC 1399 and (c,d) NGC 4874. 
All the GC candidates marked in Figure \ref{fig_gcselect2} are also marked in pink dots.
The bright GC candidates in NGC 1399 ($19.5<I_{814,0}<23.5$) and NGC 4874 ($23.0<I_{814,0}<25.0$)
are 
marked in black circles, 
and the corresponding magnitude ranges are marked in dotted lines.
The solid lines denote the color range for selecting the GCs 
and the dashed lines denote the color used for dividing blue and red GC population. 
}
\label{fig_n4874cmd}
\end{figure*}



Before estimating the fraction of red GCs in each MCEG, 
we first examine the color distributions of the GCs in NGC 1399 and NGC 4874 for references. 
Previous studies of extragalactic GCs based on F814W/F160W photometry 
are 
performed for only a few galaxies in the literature. 
To date, NGC 1399 \citep{bla12} and NGC 4874 \citep{cho16} are 
the only galaxies with those colors measured so they are 
good references 
of massive ETGs for this study. 
NGC 1399 is a cD galaxy in the Fornax cluster with 
$R_e=10.7$ kpc and $M_* \approx 2.6\times10^{11} M_\sun$ \citep{liu19}.
NGC 4874 is a cD galaxy in the Coma cluster with 
$R_e=22.7$ kpc and $M_* \approx 5.8 \times10^{11} M_\sun$ \citep{wei14}. 
We adopt the effective radius and the stellar mass of NGC 1399/NGC 4874 
from the ACS Fornax Cluster Survey (ACSFCS)/ACS Coma Cluster Survey (ACSCCS).
The distances adopted from those surveys are 20 Mpc \citep[$(m-M)_0=31.5$;][]{bla09} for NGC 1399 and 100 Mpc \citep[$(m-M)_0=35.0$;][]{car08} for NGC 4874. 
The properties of these galaxies are also summarized in Table \ref{tab_list}.

There are several 
reasons 
for referencing these 
studies for NGC 1399 and NGC 4874.
First, they use the same F814W($I_{814}$)/F160W($H_{160}$) filter systems as our study of MCEGs. 
This serves as useful references to compare GC systems between MCEGs and massive ETGs. 
We can set the selection criteria for GCs in $(I_{814}-H_{160})$ colors. 
Second, they additionally use the F475W($g_{475}$) filter system. 
We can set the criteria for dividing blue and red GC subpopulations in $(g_{475}-I_{814})$ colors. 
Third, the distances to our MCEGs are in between the distances to NGC 1399 and NGC 4874. 
Therefore, the selection criteria for GCs and 
the color division criteria 
determined from NGC 1399 and NGC 4874 can be similarly applied to the MCEGs.

The most important reason we need these references is that $(I_{814}-H_{160})$ colors are not effective to divide GC samples into two separate subpopulations. 
\citet{bla12} presented $(g_{475}-I_{814})$, $(g_{475}-z_{850})$ and $(I_{814}-H_{160})$ colors 
of the GCs in NGC 1399, and showed that 
the bimodality is clearly seen in both $(g_{475}-I_{814})$ and $(g_{475}-z_{850})$ color distributions 
while it is less clear in the $(I_{814}-H_{160})$ color distribution.
\citet{cho16} presented similar results 
for NGC 4874.


Unlike $(I_{814}-H_{160})$ colors, $(g_{475}-I_{814})$ colors are effective to divide GCs into two subpopulations. 
Mixture modelling algorithms such as KMM \citep{ash94} or Gaussian Mixture Modeling \citep[GMM;][]{mur10} 
is often applied to optical color distributions
to derive the fractions of blue and red subpopulations \citep[e.g.][]{pen06,bla12,lee19}. 
Therefore, from $g_{475},I_{814},H_{160}$ photometric studies of GCs in NGC 1399 and NGC 4874,
we first determine a $(g_{475}-I_{814})$ color criterion for dividing blue and red subpopulations. 
Then we 
determine a 
corresponding $(I_{814}-H_{160})$ color criterion 
from the relations between those two colors 
derived in this study. More details are explained in the next section. 
We use this fixed $(I_{814}-H_{160})$ color criterion to estimate the
GC fractions of each MCEG. 
Using cD galaxies as a reference does not bias the color-color relation because the intrinsic properties of GC subpopulations (age, metallicity, etc.) are similar in every massive galaxy \citep{bro06}. 

\begin{figure*}
\centering
\includegraphics[scale=0.65]{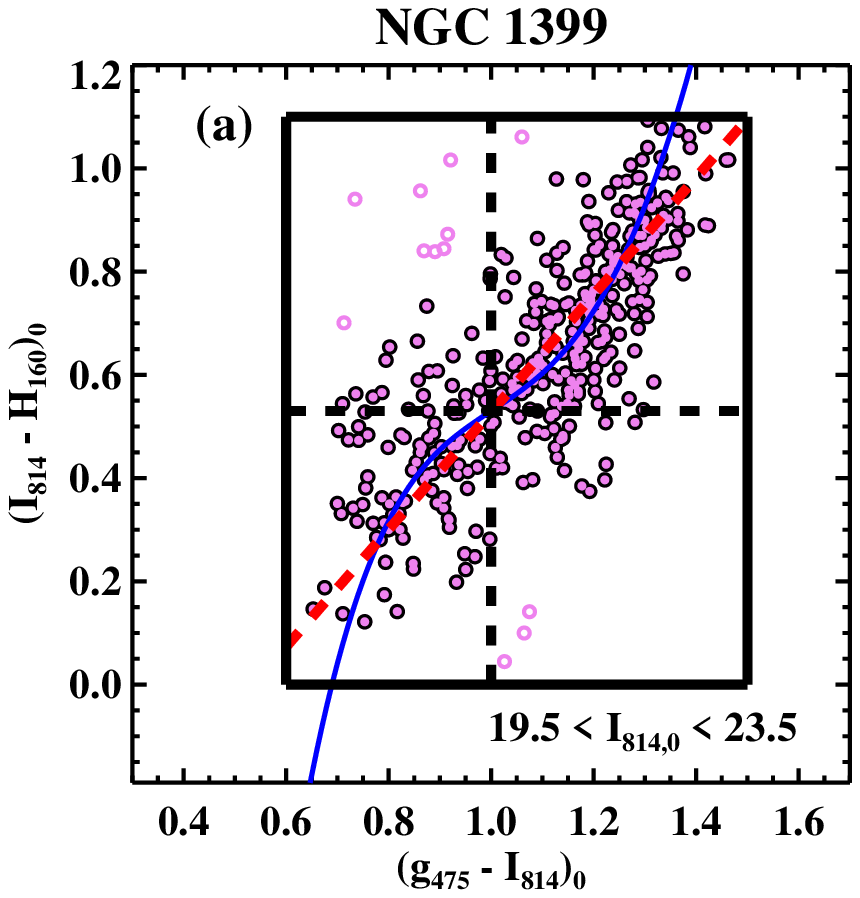}
\includegraphics[scale=0.65]{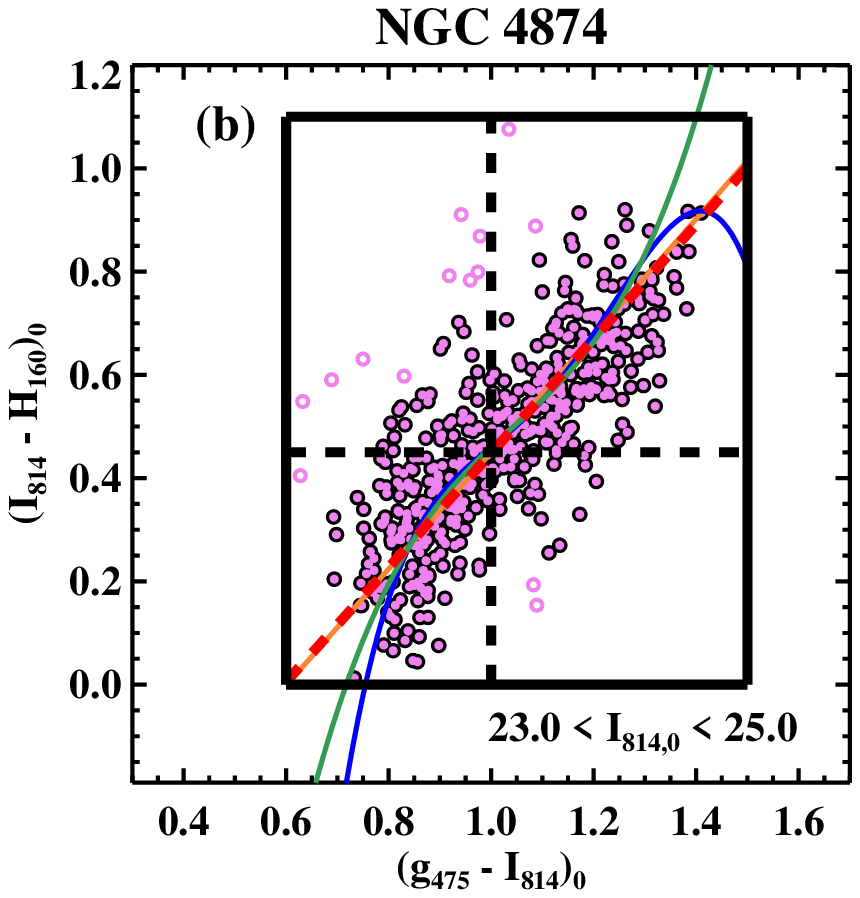}
\includegraphics[scale=0.65]{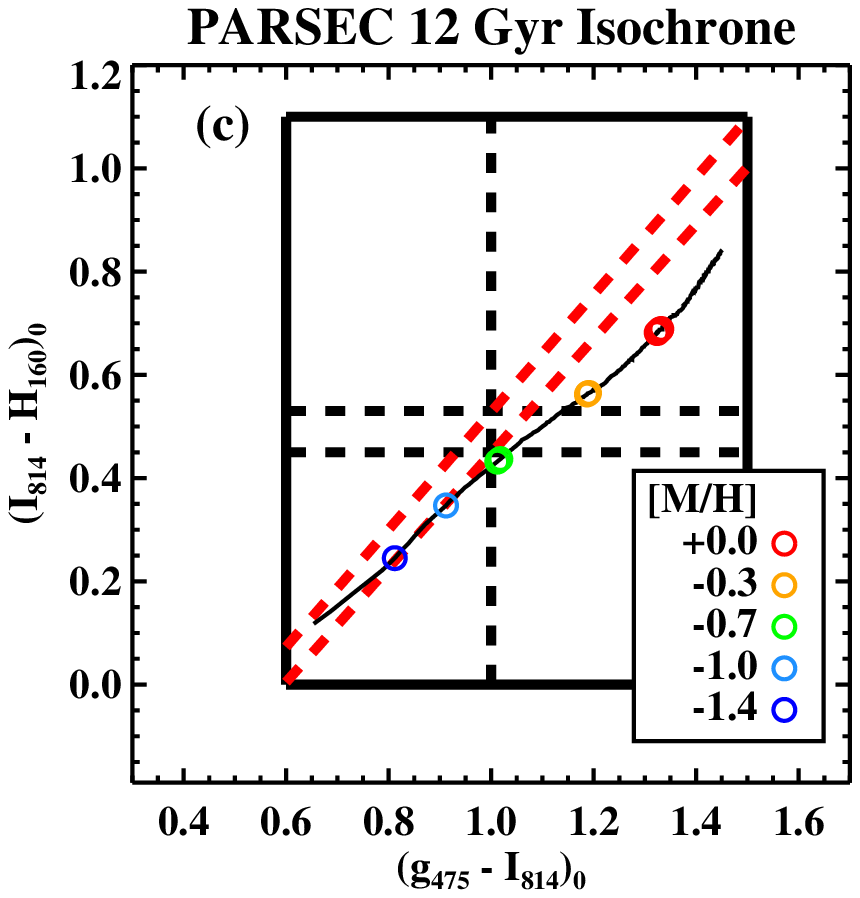}
\caption{
Color-color diagrams of the bright GC candidates in (a) NGC 1399 and (b) NGC 4874. 
The blue line in 
(a) is a quartic relation for NGC 1399 GCs given by \citet{bla12} 
shifted by $-0.04$ in y-axis.
The orange, green, and blue lines in 
(b) are linear, cubic, and quartic relations for NGC 4874 GCs given by \citet{cho16}. 
The red dashed lines in (a,b) are linear relations derived in this study for each galaxy after 3$\sigma$ clipping.
The clipped sources are marked in open pink circles, 
and filled pink circles are sources used to derive the relation. 
(c) Color-color relation derived from PARSEC 12 Gyr isochrone. We mark [M/H] values along with the relation. 
The same red dashed lines in (a,b) are also marked for comparison. 
In all the three diagrams, 
the black solid lines denote the color range for selecting the GCs 
and the dashed lines denote the color used for dividing blue and red GC populations. 
}
\label{fig_n4874ccd}
\end{figure*}

\begin{figure}
\centering
\includegraphics[scale=0.8]{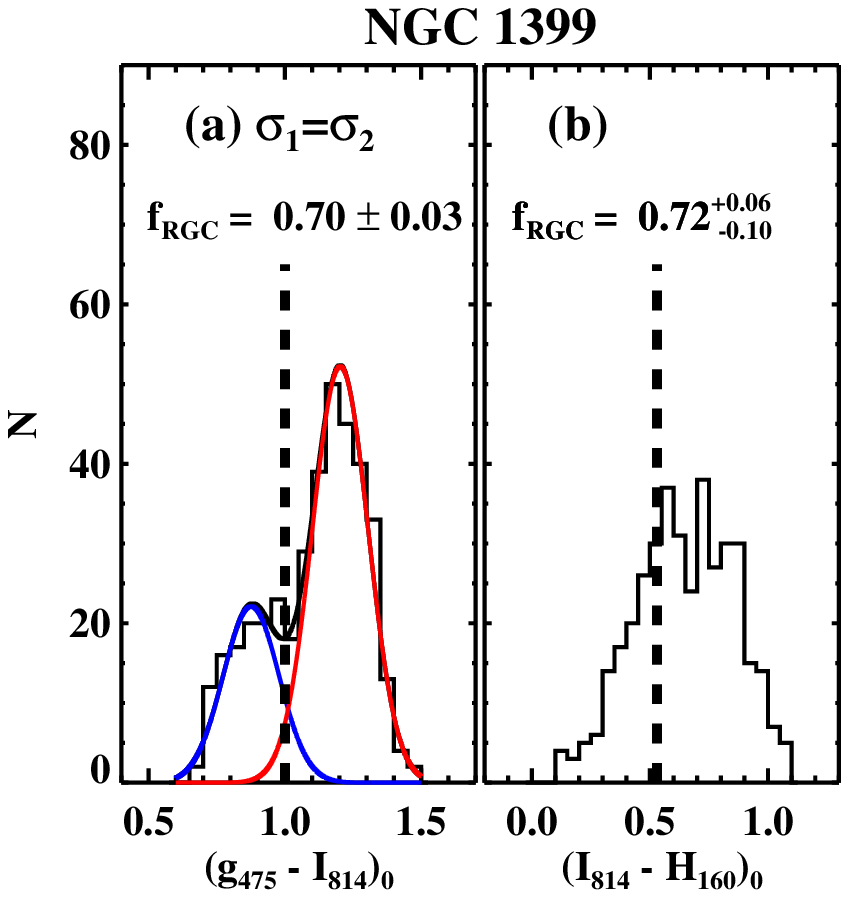}
\includegraphics[scale=0.8]{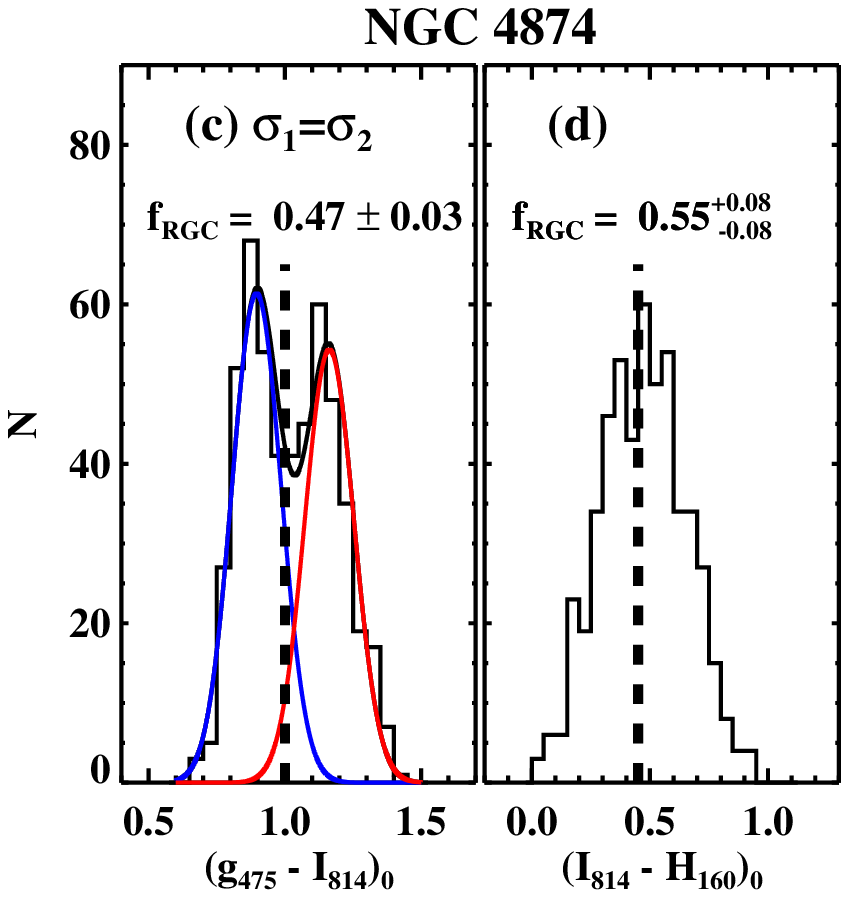}
\caption{
Color distributions of the bright GCs in (a,b) NGC 1399 ($19.5<I_{814,0}<23.5$) 
and (c,d) NGC 4874 ($23.0<I_{814,0}<25.0$). 
GMM analysis results for $(g_{475}-I_{814})_0$ colors with an equal width option 
are marked in blue and red curves. 
The dashed lines denote the color used for dividing blue and red GC populations. 
}
\label{fig_n4874cdfhomo}
\end{figure}

\begin{figure}
\centering
\includegraphics[scale=0.8]{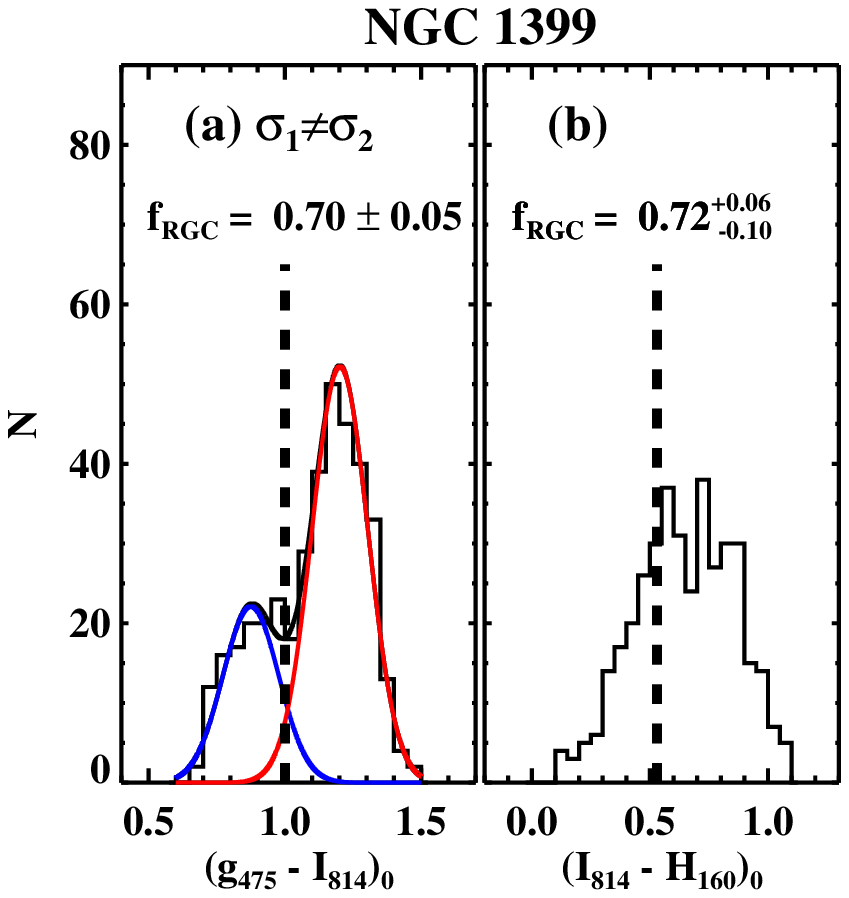}
\includegraphics[scale=0.8]{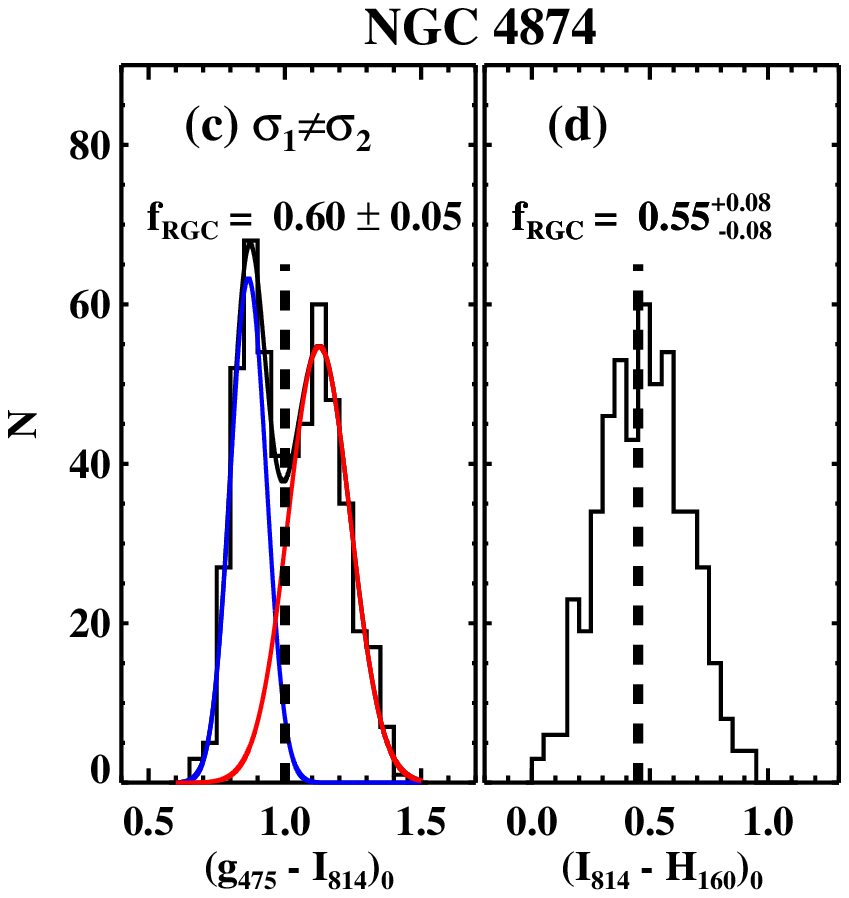}
\caption{
Same plot as Figure \ref{fig_n4874cdfhomo} but the GMM analysis 
results with an unequal width option are marked.
}
\label{fig_n4874cdfhetero}
\end{figure}

\subsection{Color Distributions of the GCs in NGC 1399 and NGC 4874} \label{n4874}

Figure \ref{fig_n4874cmd} shows $I_{814,0}$ vs. $(g_{475}-I_{814})_0$ and $I_{814,0}$ vs. $(I_{814}-H_{160})_0$ CMDs
of the GC candidates in NGC 1399 and NGC 4874 selected in the previous section.
Note that most of the GC candidates selected 
through size information exhibit GC-like colors, 
meaning that they are indeed GC candidates.
We set the color range to select GCs in each galaxy to be 
$0.6<(g_{475}-I_{814})_0<1.5$ and $0.0<(I_{814}-H_{160})_0<1.1$ based on the bright GC candidates 
in NGC 1399 ($19.5<I_{814,0}<23.5$ mag) and NGC 4874 ($23.0<I_{814,0}<25.0$ mag).
The magnitude ranges of the bright GC candidates are adopted from the previous studies of each galaxy 
which were used to derive the color-color relation.
We follow those magnitude ranges for the bright GCs hereafter.

Figures \ref{fig_n4874ccd}(a) and (b) shows $(I_{814}-H_{160})_0$ vs. $(g_{475}-I_{814})_0$ color-color diagrams 
of the bright GCs in NGC 1399 and NGC 4874.
We overlay linear color-color relations derived in this study after 3$\sigma$ clipping iteratively (red dashed lines): 
$(I_{814}-H_{160})_0=(1.14\pm0.03)(g_{475}-I_{814})_0 - (0.61\pm0.04)$ with an RMS value of 0.13 
for NGC 1399 and 
$(I_{814}-H_{160})_0=(1.11\pm0.03)(g_{475}-I_{814})_0 - (0.66\pm0.03)$ with an RMS value of 0.11
for NGC 4874. 
If we change the faint magnitude limit to include fainter GCs, 
the scatter of the relation increases but the overall relation changes little. 
The color-color relations suggested by \citet{bla12} (eq. 1) and \citet{cho16} (eqs. 9 to 11) are also overlaid.
In the case of NGC 4874, the bright GCs follow the relations very well 
and the linear relation of the GCs in this study 
shows a tight correlation with the relations of \citet{cho16}. 
This means that the photometric results derived in this study agree well with the previous study.
However, in the case of NGC 1399, 
there is a slight disagreement between our relation and the previously derived relation.
This difference 
arises because \citet{bla12} did not consider color corrections caused by different apertures.
In Section \ref{apcor}, we correct $(g_{475}-I_{814})$ color as much as $2.5{\rm log}(0.794/0.770)=0.033$
and $(I_{814}-H_{160})$ color as much as $2.5{\rm log}(0.770/0.789)=-0.026$, 
which can explain the difference between our relation and \citet{bla12}'s relation. 
Therefore, we mark the relation after shifting the relation by -0.04 in y-axis to be consistent with the fitting of this study.

\begin{deluxetable*}{llllllllllll}
\tabletypesize{\footnotesize}
\tablecaption{The GMM Analysis Results for the $(g_{475}-I_{814})_0$ color distributions of the GCs in NGC 1399 and NGC 4874 
\label{tab_gmm} }
\tablewidth{0pt}
\tablehead{
\colhead{Galaxy} & \colhead{Case} & \colhead{$\mu_1$} & \colhead{$\sigma_1$} & \colhead{$N_1$} & 
\colhead{$\mu_2$} & \colhead{$\sigma_2$} & \colhead{$N_2$} & \colhead{$f_2$} & \colhead{$D$} & \colhead{$p$} & \colhead{$k$}
}
\startdata
NGC 1399 & $\sigma_1=\sigma_2$   & $0.88 \pm 0.01$ & $0.10 \pm 0.01$ & $115 \pm 11$ & $1.20 \pm 0.01$ & $0.10 \pm 0.01$ & $268 \pm 11$ & $0.70 \pm 0.03$ & $3.2 \pm 0.2$ & $<10^{-3}$ & $-0.681$ \\
         & $\sigma_1\ne\sigma_2$ & $0.88 \pm 0.03$ & $0.10 \pm 0.01$ & $116 \pm 19$ & $1.20 \pm 0.01$ & $0.10 \pm 0.01$ & $267 \pm 19$ & $0.70 \pm 0.05$ & $3.2 \pm 0.3$ & $<10^{-3}$ & $-0.681$ \\
\hline
NGC 4874 & $\sigma_1=\sigma_2$ & $0.90 \pm 0.01$ & $0.09 \pm 0.01$ & $278 \pm 14$ & $1.16 \pm 0.01$ & $0.09 \pm 0.01$ & $245 \pm 14$ & $0.47 \pm 0.03$ & $3.0 \pm 0.2$ & $<10^{-3}$ & $-0.966$ \\
         & $\sigma_1\ne\sigma_2$ & $0.87 \pm 0.01$ & $0.07 \pm 0.01$ & $212 \pm 24$ & $1.13 \pm 0.01$ & $0.11 \pm 0.01$ & $311 \pm 24$ & $0.60 \pm 0.05$ & $2.8 \pm 0.2$ & $<10^{-3}$ & $-0.966$
\enddata
\tablecomments{See Figures \ref{fig_n4874cdfhomo} and \ref{fig_n4874cdfhetero}.} 
\end{deluxetable*}

Another feature is that $(I_{814}-H_{160})_0$ colors 
for given $(g_{475}-I_{814})$ colors of the NGC 4874 GCs are, on average, 
0.1 bluer than those of the NGC 1399 GCs. 
This color difference also exists between the relation suggested by \citet{bla12} and \citet{cho16}, as already pointed out in \citet{cho16}.
This is caused mainly by the difference in the distances to the two galaxies. 
The distance to NGC 1399 is about five times smaller than the distance to NGC 4874 
so that the light distribution of the GCs in NGC 1399 deviate from that of the PSF 
more than that of the GCs in NGC 4874.
This difference appears mostly in $(I_{814}-H_{160})$ color rather than in $(g_{475}-I_{814})$ color.
In this study we apply the same aperture correction to both galaxies 
and adopt the different color-color relation of each galaxy.

Figure \ref{fig_n4874ccd}(c) shows a color-color relation derived from PAdova and TRieste Stellar Evolution Code \citep[PARSEC v1.2S;][]{bre12} 12 Gyr isochrone. The simple stellar population integrated magnitudes are obtained with the default options in the web interface CMD 3.1\footnote{\url{http://stev.oapd.inaf.it/cgi-bin/cmd_3.1}}, e.g. Chabrier lognormal IMF. Note that 
the division color at 
$(g_{475}-I_{814})=1.0$ in the model corresponds to $(I_{814}-H_{160})=0.42$, which is very similar to the case for NGC 4874 ($(I_{814}-H_{160})=0.45$) 
and is 0.11 smaller than the value for NGC 1399 ($(I_{814}-H_{160})=0.53$). 

Figures \ref{fig_n4874cdfhomo} and \ref{fig_n4874cdfhetero} show 
the color distributions of the bright GCs in NGC 1399 and NGC 4874. 
In the $(g_{475}-I_{814})_0$ distribution of the GCs in NGC 1399, 
a minimum between two peaks (a dip) is found to be at $(g_{475}-I_{814})_0\approx1.0$ 
\citep[see also Figure 8 in][]{bla12}.
Similarly, a dip is seen clearly at $(g_{475}-I_{814})_0\approx1.0$ for NGC 4874 GCs 
\citep[see also Figure 13 in][]{cho16}. 
These results are very similar to the results from the previous studies.

To divide 
the GCs into two subpopulations statistically, 
we apply GMM analysis to $(g_{475}-I_{814})_0$ distribution of the GCs \citep{mur10}.
Figure \ref{fig_n4874cdfhomo} is based on the GMM analysis
with an equal variance option (i.e. homoscedastic case)
and Figure \ref{fig_n4874cdfhetero} with an unequal variance option (i.e. heteroscedastic case).
The results are summarized in Table \ref{tab_gmm}. 
According to the analysis, there is a clear bimodality in $(g_{475}-I_{814})_0$ color 
for both homoscedastic and heteroscedastic cases (separation $D>2$, p-value $p<10^{-3}$, and kurtosis $k<0$).
Moreover, the two populations are divided almost at the same color, $(g_{475}-I_{814})_0\approx1.0$, for all cases.
By applying different color-color relation for each galaxy, 
$(g_{475}-I_{814})_0=1.0$ corresponds to 
$(I_{814}-H_{160})_0=0.53\pm0.05$ for NGC 1399 
and $0.45\pm0.04$ for NGC 4874. 
These values are also marked in Figures \ref{fig_n4874cmd} and \ref{fig_n4874ccd} 
and are used to divide the
GCs in our target MCEGs.

\subsection{Estimation of the Red GC Fractions of NGC 1399 and NGC 4874}

We 
estimate the fraction of red GCs in NGC 1399 and NGC 4874 using the photometry derived in this study.
In the case of $(g_{475}-I_{814})_0$ color distributions for NGC 1399, 
we obtain 
$f_{RGC}=0.70\pm0.03$ if we apply the GMM with an equal variance option
and $0.70\pm0.05$ if we adopt an unequal variance option.
This is consistent with the results of \citet{bla12}, $0.68\pm0.04$ and $0.71\pm0.03$. 
In the case of $(g_{475}-I_{814})_0$ color distributions for NGC 4874, 
we obtain 
$f_{RGC}=0.47\pm0.03$ if we apply the GMM with an equal variance option 
and 
$0.60\pm0.05$ if we adopt an unequal variance option.
This is also consistent with the results of \citet{cho16}, $f_{RGC}=0.493\pm0.031$ and $0.604\pm0.083$. 
From $(I_{814}-H_{160})_0$ color distributions with a fixed division color, 
we derive $f_{RGC}=0.72^{+0.06}_{-0.10}$ for NGC 1399 and $0.55^{+0.08}_{-0.08}$ for NGC 4874. 
The errors for the red GC fraction are determined from 
the error for the division color. 


The most important result from the analysis of NGC 1399 and NGC 4874 GCs is that 
the values of red GC fraction derived from $(g_{475}-I_{814})_0$ and $(I_{814}-H_{160})_0$ are very similar.
This indicates that the fractions of red GCs derived from 
the two colors are consistent with each other. 
Therefore, we can safely derive the red GC fraction based on $(I_{814}-H_{160})_0$ color for our target MCEGs.
Moreover, the distances to our target MCEGs are mostly in between the distance to NGC 1399 and NGC 4874, 
so blue and red GCs would be divided at $(I_{814}-H_{160})_0=0.45$ to 0.53.
In the next section, we adopt both $(I_{814}-H_{160})_0=0.45$ and 0.53 
and derive the minimum and the maximum value of the red GC fractions.
The mean red GC fractions are derived using  $(I_{814}-H_{160})_0=0.49\pm0.04$, an average value of 0.45 and 0.53.

\begin{figure*}
\centering
\includegraphics[scale=0.8]{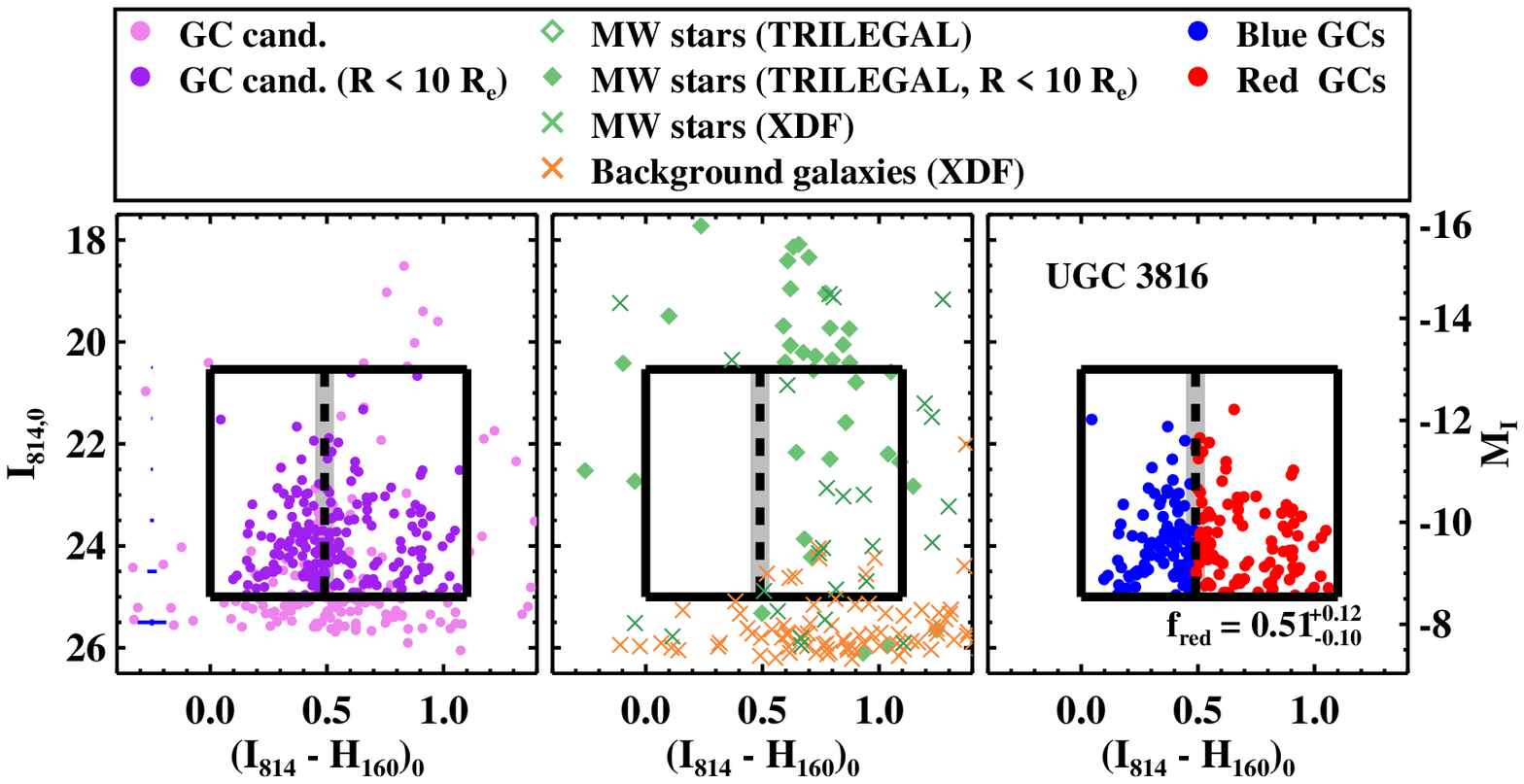} 
\includegraphics[scale=0.8]{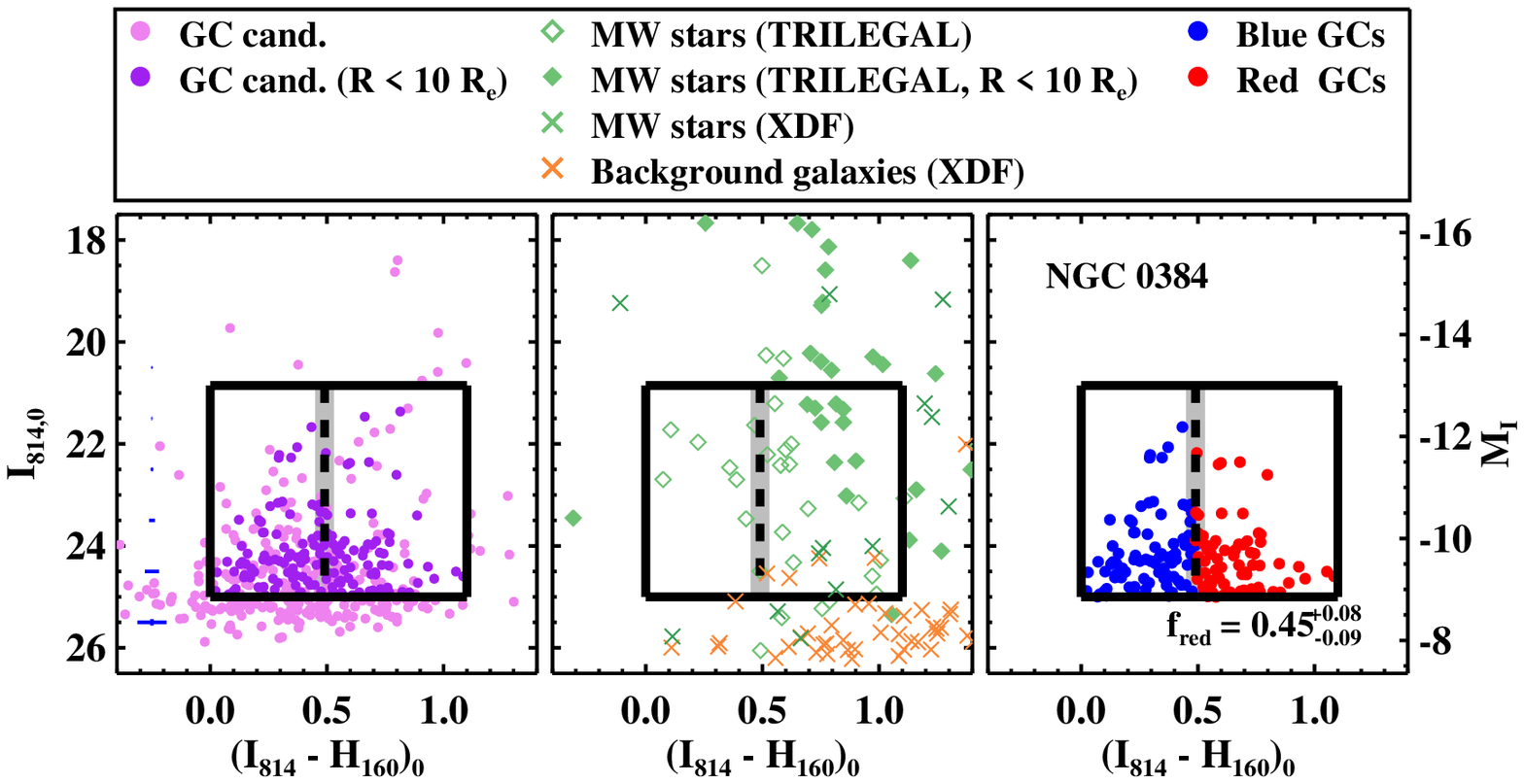} 
\caption{
(Left) CMDs of the GC candidates of our 
12 target MCEGs. 
All the GC candidates marked in Figure \ref{fig_gcselect} are also marked in pink dots.
The selected GCs with $R<10R_{e,circ}$ within the black boxes are marked in purple dots.
The color range of the GCs is $0.0<(I_{814}-H_{160})_0<1.1$ 
and the magnitude range is $M_I>-13.0$ and $I_{814,0}<25.0$. 
Blue errorbars in the left denote mean photometric errors for given magnitude bins. 
(Middle) CMDs of the MW stars generated by TRILEGAL model (green diamonds) 
and foreground/background sources found in XDF (green and orange crosses).
(Right) CMDs of the GC candidates after removing foreground stars with $R<10R_{e,circ}$ statistically.
We mark blue and red GCs according to the color division $(I_{814}-H_{160})_0=0.49\pm0.04$ as marked in dashed lines,  
and mark the red GC fractions. 
}
\label{fig_cmd01}
\end{figure*}

\begin{figure*}
\centering
\includegraphics[scale=0.8]{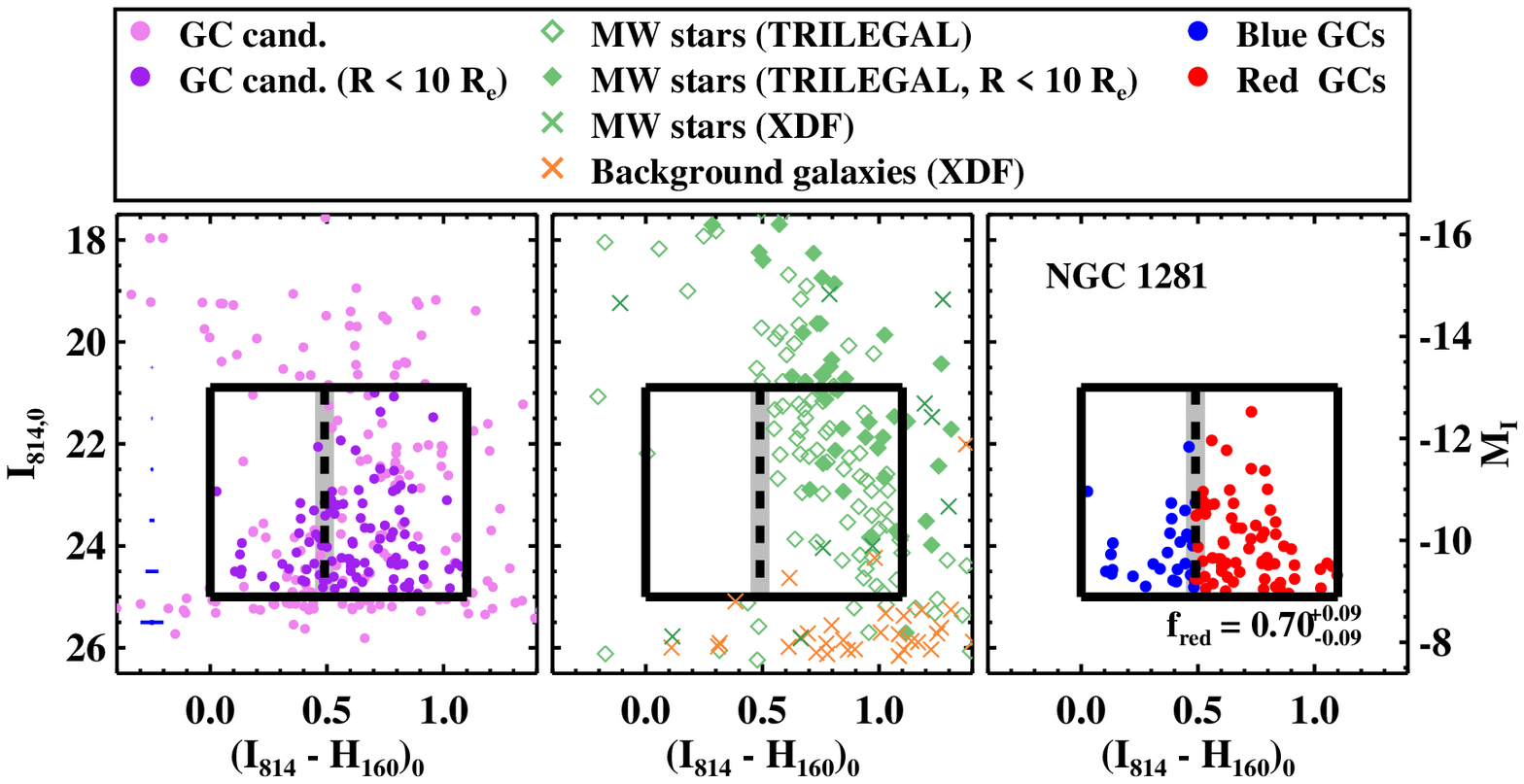} 
\includegraphics[scale=0.8]{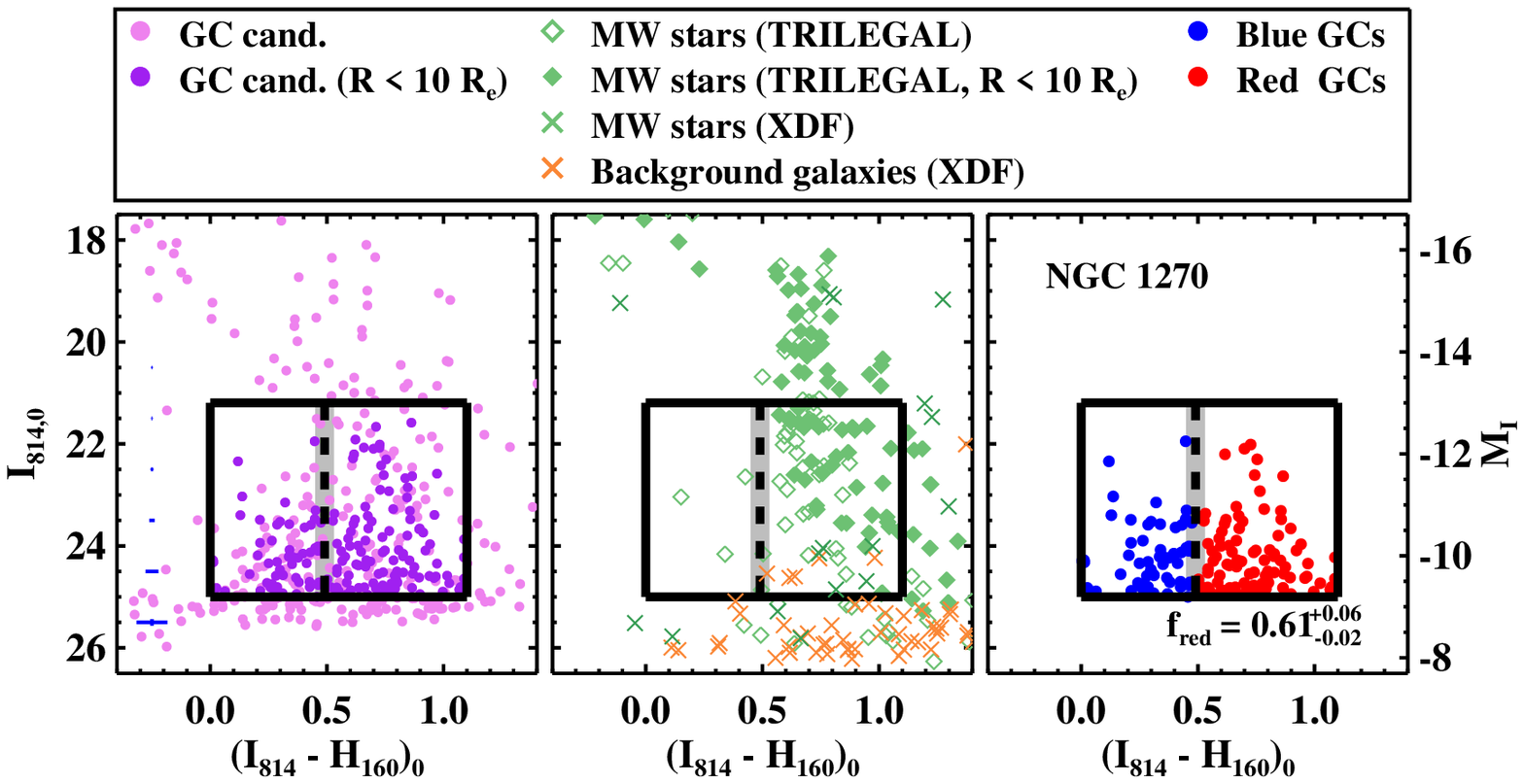} 
\caption{Figure \ref{fig_cmd01} continued.
}
\label{fig_cmd03}
\end{figure*}


\begin{figure*}
\centering
\includegraphics[scale=0.8]{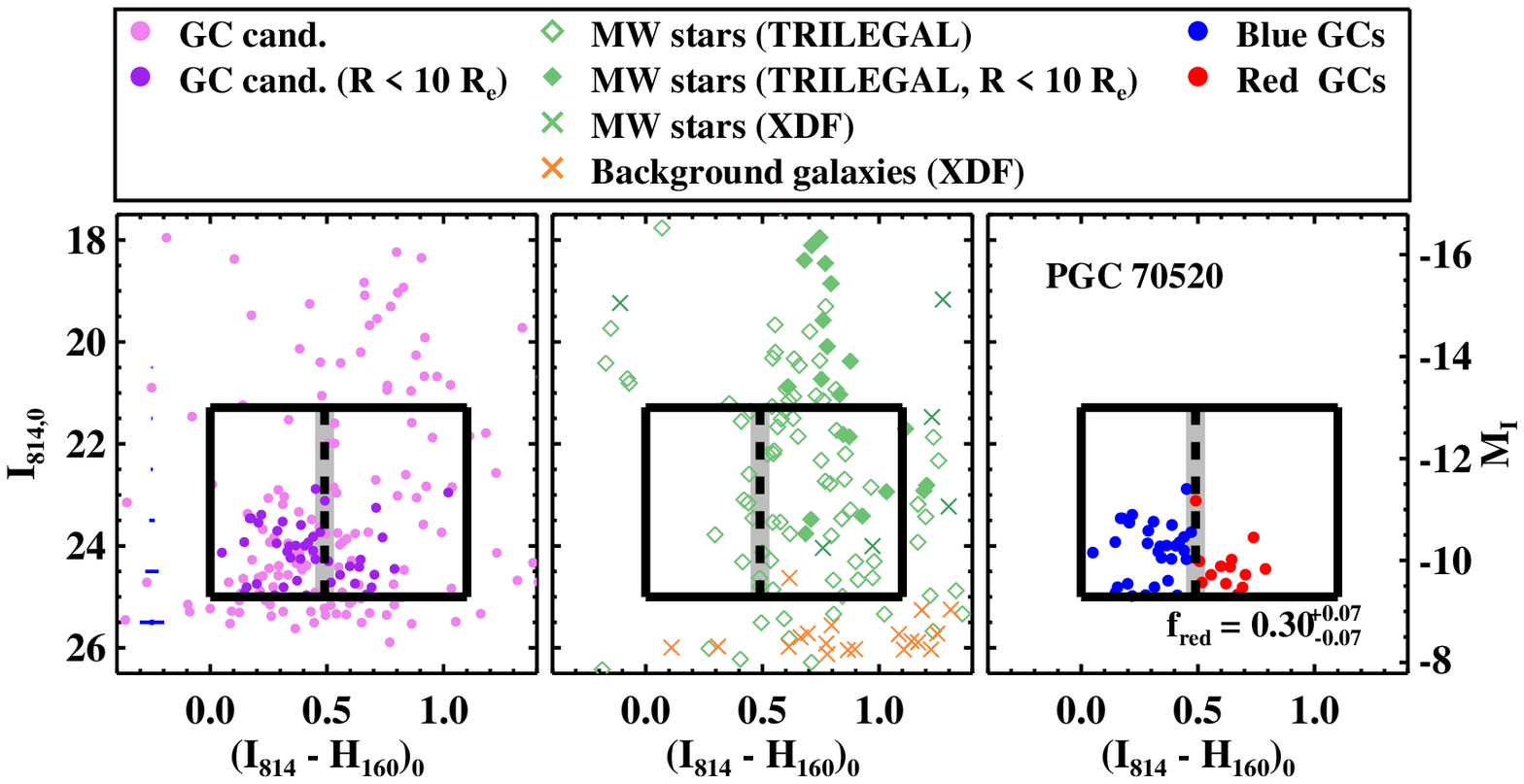} 
\includegraphics[scale=0.8]{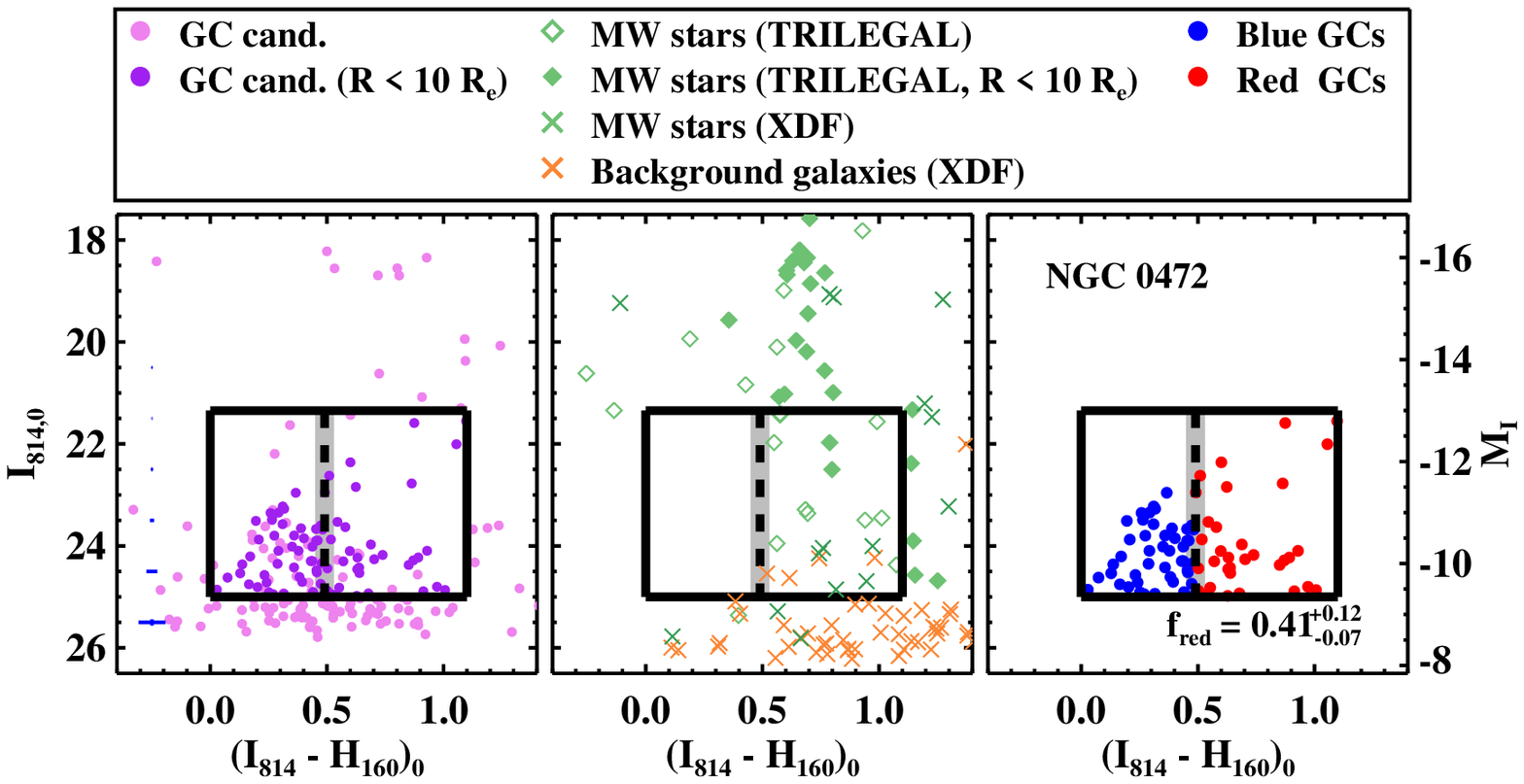} 
\caption{Figure \ref{fig_cmd01} continued.
}
\label{fig_cmd07}
\end{figure*}

\begin{figure*}
\centering
\includegraphics[scale=0.8]{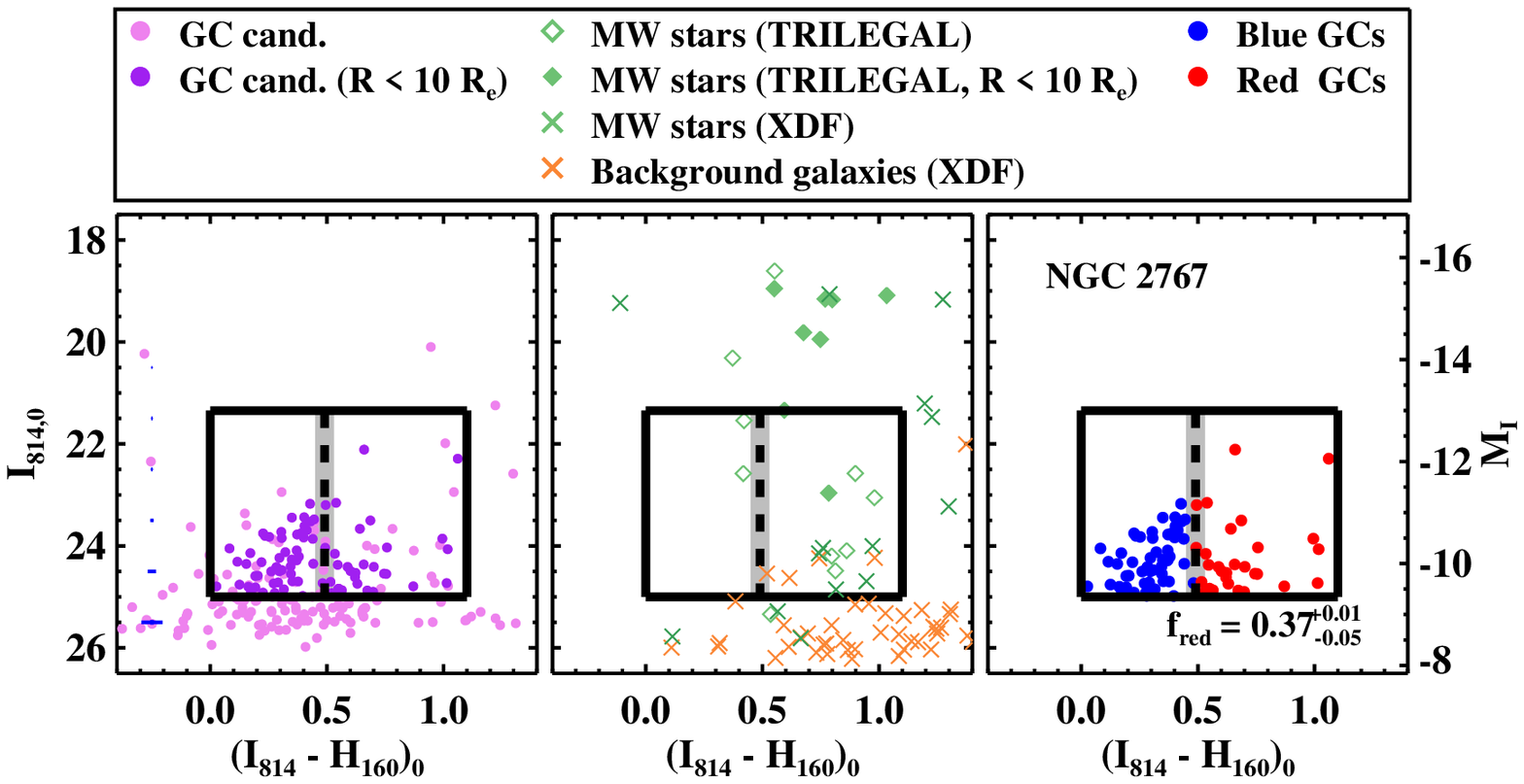} 
\includegraphics[scale=0.8]{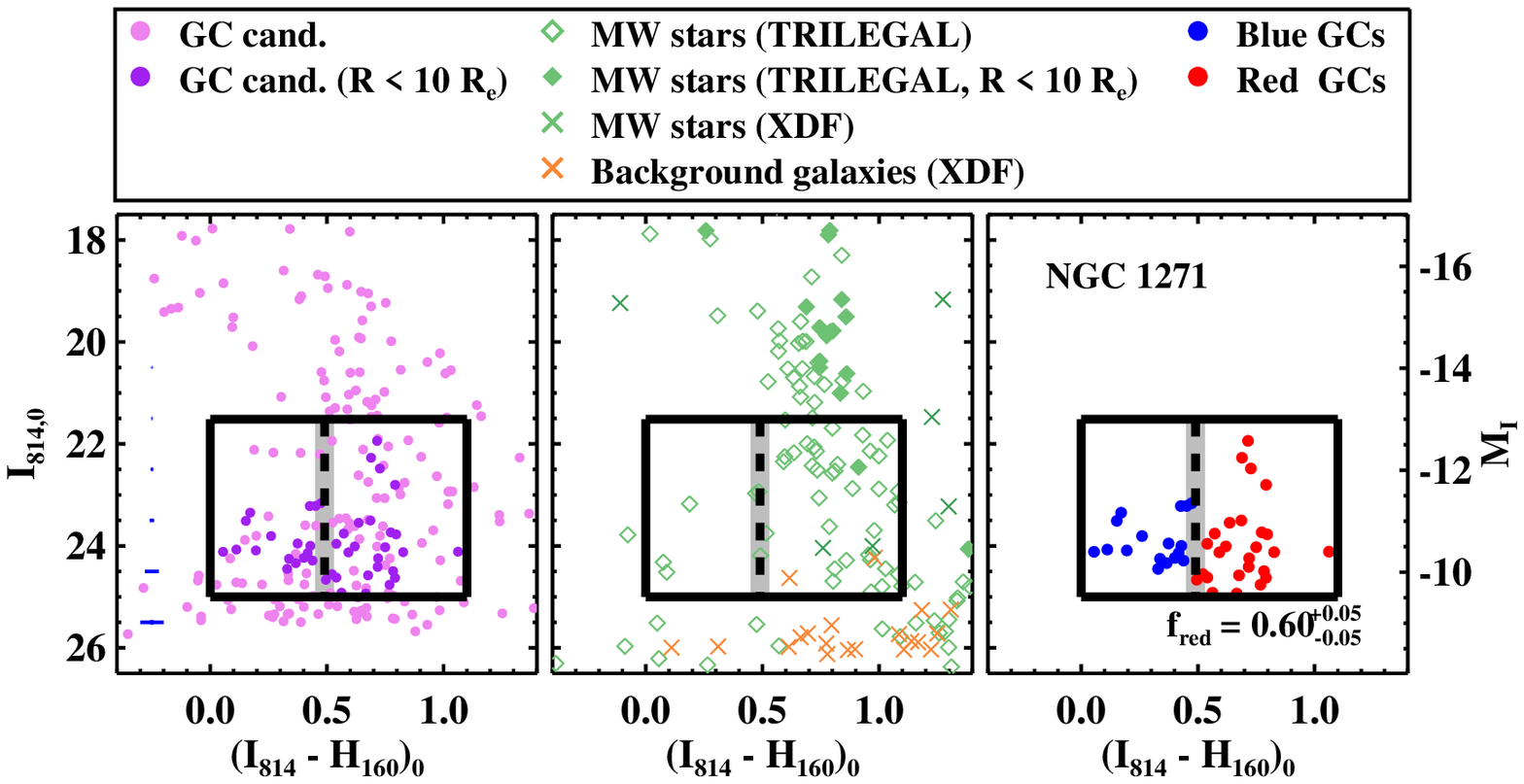} 
\caption{Figure \ref{fig_cmd01} continued.
}
\label{fig_cmd09}
\end{figure*}

\begin{figure*}
\centering
\includegraphics[scale=0.8]{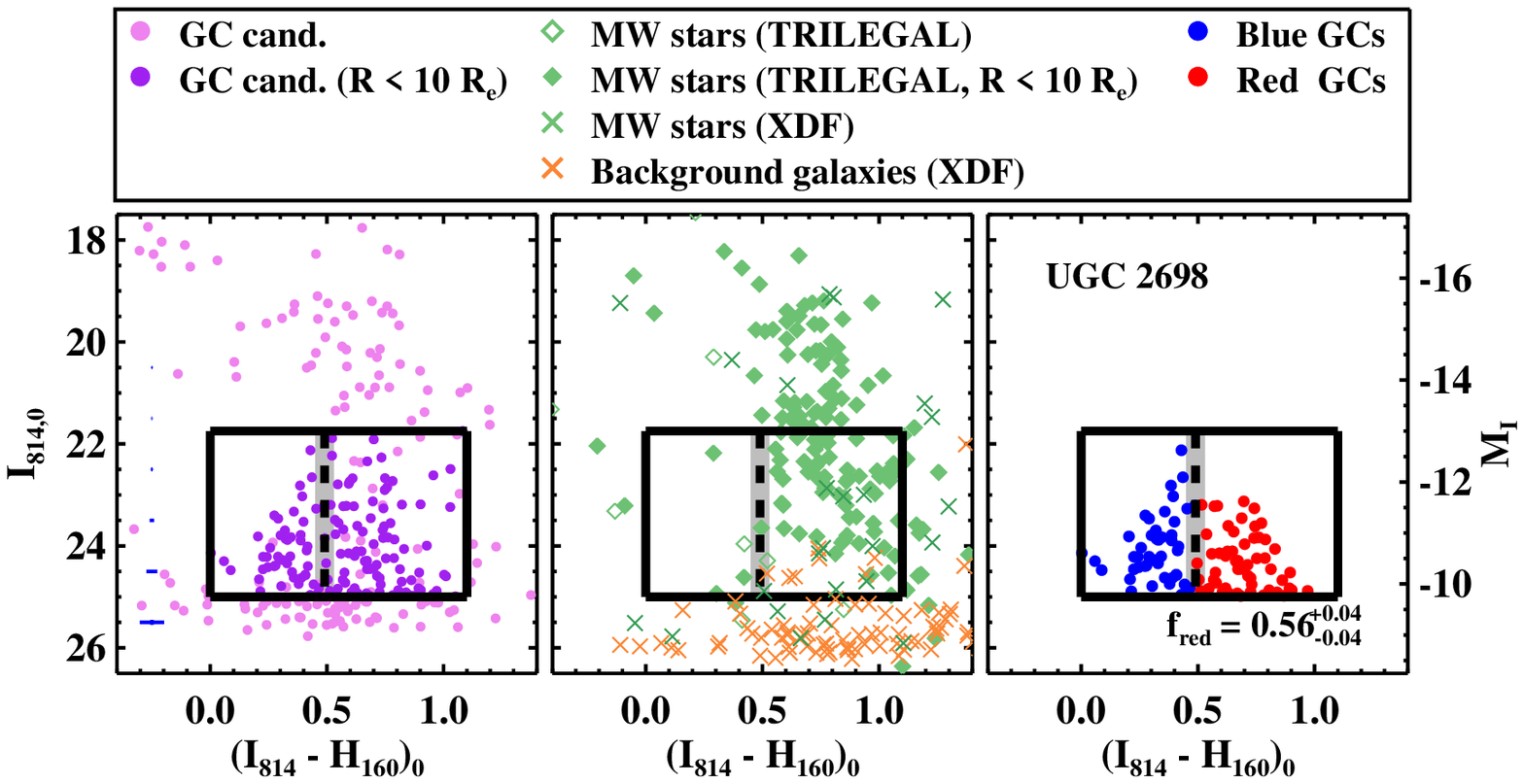} 
\includegraphics[scale=0.8]{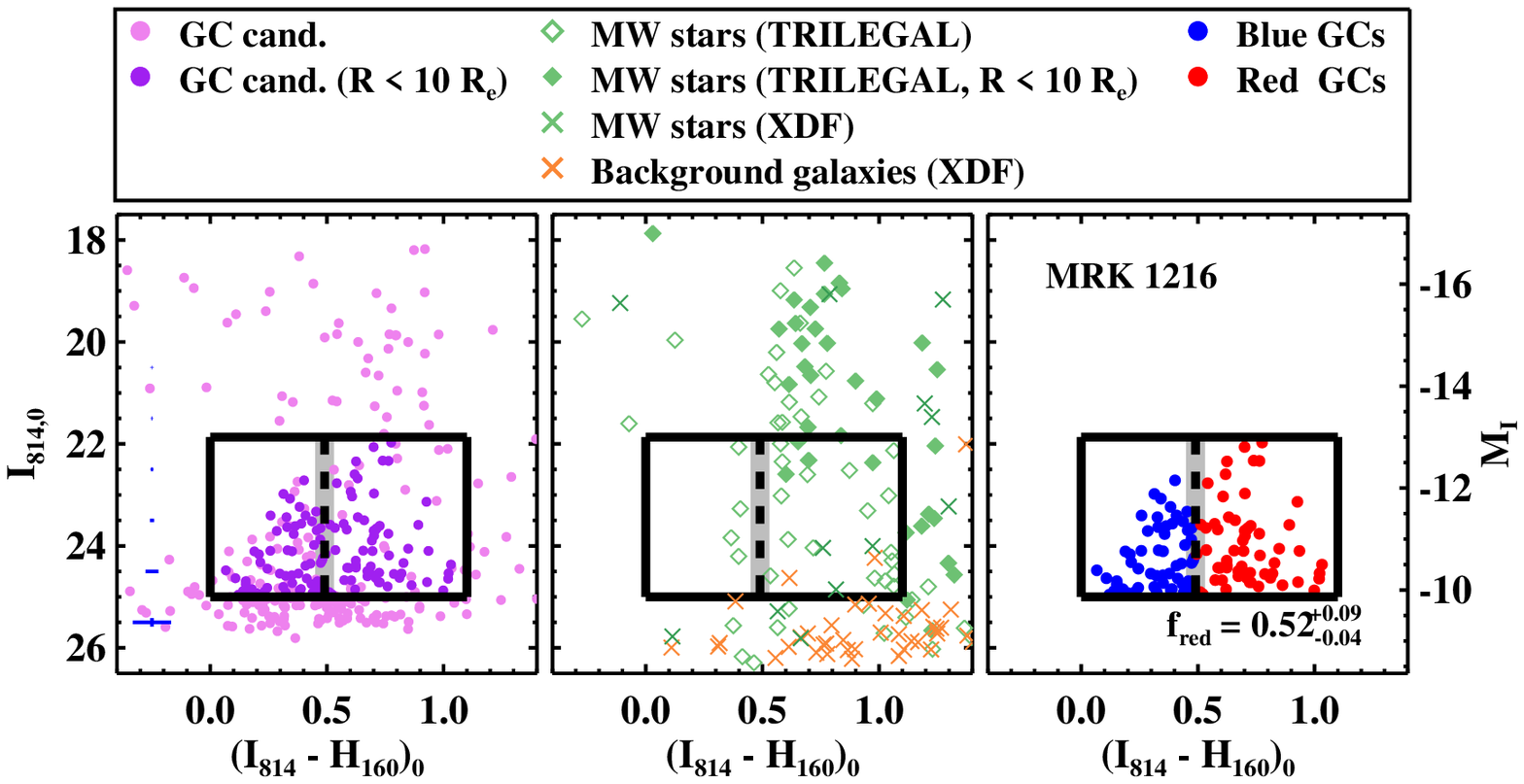} 
\caption{Figure \ref{fig_cmd01} continued.
}
\label{fig_cmd11}
\end{figure*}

\begin{figure*}
\centering
\includegraphics[scale=0.8]{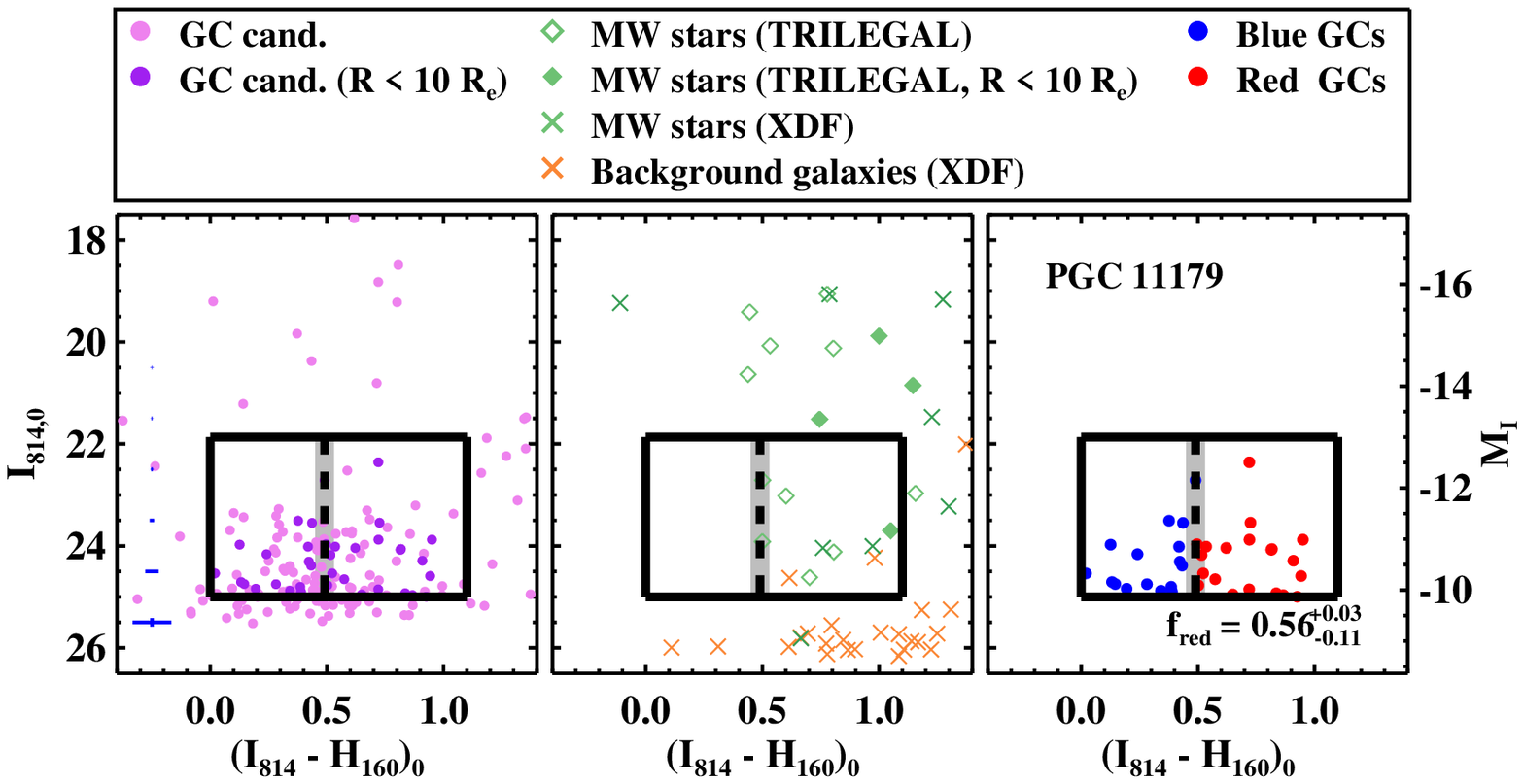} 
\includegraphics[scale=0.8]{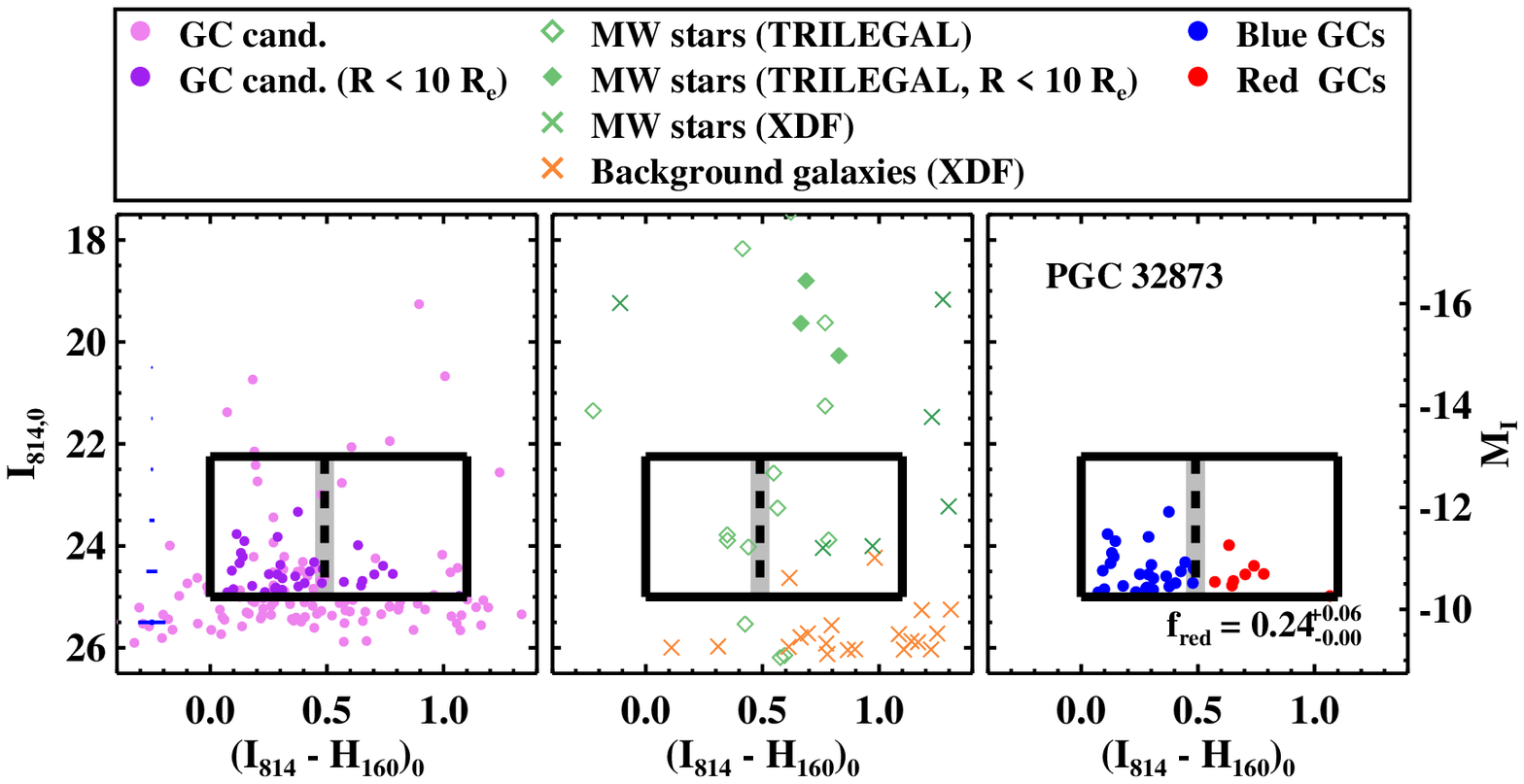} 
\caption{Figure \ref{fig_cmd01} continued.
}
\label{fig_cmd13}
\end{figure*}

\subsection{Color Distribution of the GCs in the Local MCEGs}

In the left panels of Figures \ref{fig_cmd01} to \ref{fig_cmd13} 
we display the $I_{814,0}$ vs. $(I_{814}-H_{160})_0$ CMDs of the GC candidates for the target MCEGs.
We plot the magnitudes and colors of the sources corrected for the foreground extinction 
to be consistent with each other.
In the figure, we draw three vertical lines as guide lines at $(I_{814}-H_{160})_0 =0.0$, 0.49, and 1.1. 
$(I_{814}-H_{160})_0=0.0$ and 1.1 are boundaries for the GC colors, 
and $(I_{814}-H_{160})_0=0.49$ is the dividing line 
between blue and red GCs.
This is the average 
of 0.53 and 0.45, which were determined from NGC 1399 and NGC 4874. 
We also draw two horizontal lines at $M_I=-13.0$ and $I_{814,0}=25.0$ which are boundaries for the GC magnitude cut.
The absolute magnitude range for GCs is, in general, $M_I > -13$ mag. 
Therefore, we set the upper magnitude limit to $M_I = -13$ mag for GC selection considering the distance to each galaxy. 
To derive the color distributions of the GC candidates in the MCEGs, 
we choose relatively bright GCs with $I_{814,0}<25.0$ mag 
($M_I\approx -10.0$ to $-8.5$ mag depending on the distances to the target galaxies)
which are rarely affected by incompleteness of our photometry 
and have smaller photometric errors ($err(I_{814}-H_{160})<0.1$).
We exclude GC candidates brighter than $M_I=-13.0$ mag because they are mostly foreground stars. 
At first, some of them were considered as ultra-compact dwarfs (UCDs), 
but the possibility was ruled out after comparing CMDs with model CMDs of foreground stars, 
as will be shown in the next section. 
We also choose GCs 
inside the circular region of the host galaxy ($R<10R_{e,circ}$) 
to minimize contamination from background sources and other nearby galaxies. 
The boundary at $R=10R_{e,circ}$ is large enough to select most of the GCs belonging to our target galaxies, 
and \citet{bea18} also used the same boundary at $R=10R_{e,circ}$ to estimate the red GC fraction of NGC 1277. 

The most distinguishable feature in the CMDs is
that the target MCEGs host a rich population of sources with $0.0<(I_{814}-H_{160})_0<1.1$, 
a similar color range 
for the GCs in NGC 1399 and NGC 4874, 
implying that the selected GC candidates with this color range in the target MCEGs are mostly genuine GCs. 
They are indeed GCs similar to those in NGC 1399 or NGC 4874.

\subsection{Estimation of the Red GC Fractions of the Local MCEGs} \label{frgc}

Before estimating the red GC fractions of each galaxy, 
we check foreground star contamination using the TRILEGAL MW model for each galaxy \citep{gir05}\footnote{\url{http://stev.oapd.inaf.it/cgi-bin/trilegal}}. 
We generate model stars according to the galactic coordinate of each galaxy and the field area. 
The area is restricted to $R<10R_{e,circ}$ for each galaxy 
similar to the area used to select GCs. 
We plot the CMD derived from the model in the middle panels of Figures \ref{fig_cmd01} to \ref{fig_cmd13}
in green diamonds. 
Open symbols represent model stars within the HST field 
and filled symbols represent those within $R<10R_{e,circ}$ for each galaxy. 
Clearly, the galaxies located in low galactic latitude show a large number of model stars 
(
NGC 1281, 
NGC 1270, NGC 1271, UGC 2698) 
and most of the red sources show 
a color distribution similar to that of 
model foreground stars.
Therefore, we exclude model stars statistically from our CMDs to estimate the red GC fraction properly. 
We divide the CMD region with a magnitude bin of 0.5 and a color bin of 0.1 
and then subtract the number of MW model stars from the number of GC candidates. 
The final CMDs are shown in the right panels of Figures \ref{fig_cmd01} to \ref{fig_cmd13}. 

We also check background galaxy contamination from the HST image.
The field of view of our images 
is not large enough to estimate the background contribution, 
so we use the XDF field as a background field \citep{ill13}. 
The target galaxies are located in various locations in the sky 
so the XDF may not necessarily represent a background for each target galaxy. 
However, it is still useful to estimate the background contamination level. 
We apply the same photometric procedures as used for our target galaxies to the XDF F814W/F160W images 
and derive $(I_{814}-H_{160})_0$ distribution of the point sources. 
As a result, we find 
fewer than seven GC-like background sources with $I_{814,0}<25.0$ mag and $R<10R_{e,circ}$, 
which negligibly affect 
the estimation of the red GC fraction.
They are marked in orange crosses in Figures \ref{fig_cmd01} to \ref{fig_cmd13}. 
Therefore, we 
do not correct for the background galaxy contamination.

\begin{deluxetable*}{lrrrrrr}
\tabletypesize{\footnotesize}
\tablecaption{Number of GC Candidates and Contaminants \label{tab_number}}
\tablewidth{0pt}
\tablehead{
\colhead{Galaxy} & \colhead{N(GC cand.)} & \colhead{N(foreground)} &\colhead{N(background)} & \colhead{f(foreground)} & \colhead{f(background)} & \colhead{N(final GCs)} \\
\colhead{} & \colhead{[A]} & \colhead{[B]} &\colhead{[C]} & \colhead{[B]/[A] [\%]} & \colhead{[C]/[A] [\%]} & \colhead{[A]$-$[B]}
}
\startdata
UGC 3816  & 205 & 7 & 7 & 3.41 & 3.41 & 198  \\
NGC 0384  & 158 & 3 & 4 & 1.90 & 2.53 & 155  \\ 
NGC 1281  & 104 & 8 & 2 & 7.69 & 1.92 & 96  \\
NGC 1270  & 168 & 20 & 5 & 11.90 & 2.98 & 148  \\
PGC 70520 & 45 & 2 & 1 & 4.44 & 2.22 & 43  \\
NGC 0472  & 82 & 0 & 4 & 0.00 & 4.88 & 82\\
NGC 2767  & 83 & 0 & 3 & 0.00 & 3.61 & 83 \\
NGC 1271  & 43 & 0 & 1 & 0.00 & 2.33 & 43  \\
UGC 2698  & 137 & 33 & 7 & 24.09 & 5.11 & 104  \\
MRK 1216  & 120 & 2 & 2 & 1.67 & 1.67 & 118  \\
PGC 11179 & 36 & 0 & 2 & 0.00 & 5.56 & 36 \\
PGC 32873 & 34 & 0 & 1 & 0.00 & 2.94 & 34 \\
\hline
\enddata
\tablecomments{For $R<10R_{e,circ}$}
\end{deluxetable*}

In Table \ref{tab_number} we summarize the number of GC candidates before and after contamination correction, and the number of contaminants for each galaxy. The fractions of foreground contaminants range from 0 to 24\% (mostly smaller than 10\%), and those of background contaminants are less than 6\%. 
Thus, the effect of background contaminants is not impacting the results discussed in this study. 

Now we are ready to estimate the red GC fraction of the target MCEGs. 
From the final CMDs, we divide the GCs in each of our target MCEGs into blue and red subpopulations 
using $(I_{814}-H_{160})_0=0.49$ as a division criterion: 
blue GCs with $0.0<(I_{814}-H_{160})_0<0.49$ and red GCs with $0.49\leq(I_{814}-H_{160})_0<1.1$.
Then we calculate the fraction of red GCs in the entire sample of GCs 
with $I_{814,0}<25.0$ mag in each galaxy.
Similarly, we calculate the minimum and the maximum value of the fraction of red GCs
using $(I_{814}-H_{160})_0=0.53$ and 0.45, respectively.
The results are shown in the right panels of Figures \ref{fig_cmd01} to \ref{fig_cmd13} and are summarized in Table \ref{tab_list}.
The fractions of red GCs in 
our target MCEGs range from  $f_{RGC} =0.2$ to 0.7, 
and NGC 1281 shows the highest red GC fraction, $f_{RGC} =0.70\pm0.09$.

\begin{figure*}
\centering
\includegraphics[scale=0.8]{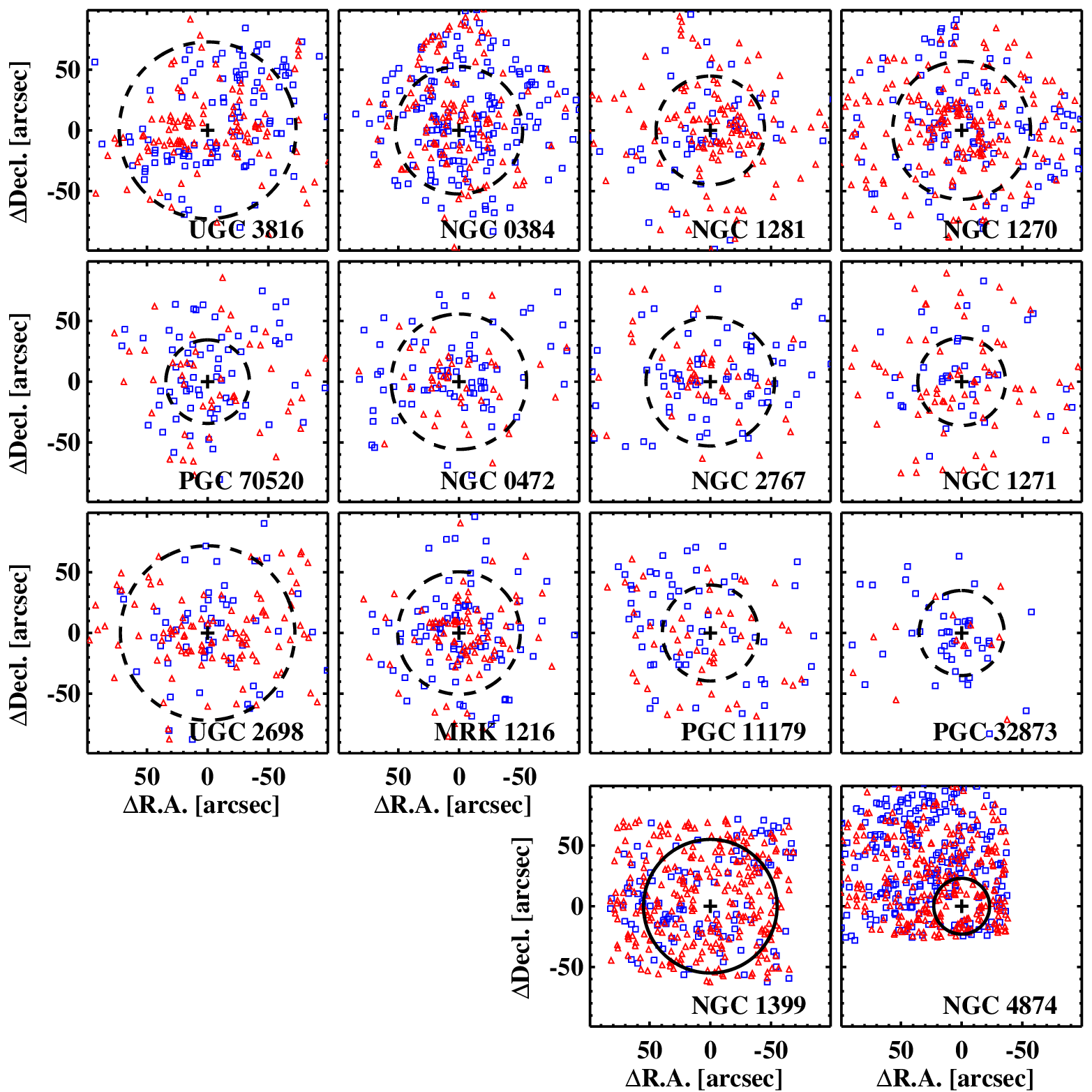}
\caption{Spatial distributions of the bright sources detected in each target MCEG and the two reference galaxies. 
Blue rectangles and red triangles mark blue and red GCs. 
Large dashed circles denote $R=10R_{e,circ}$ of each galaxy which is a boundary of the GC systems used in this study.
For the case of the two reference galaxies, large solid circles denote $R=0.5R_e$.
}
\label{fig_map1}
\end{figure*}

\begin{figure*}
\centering
\includegraphics[scale=0.8]{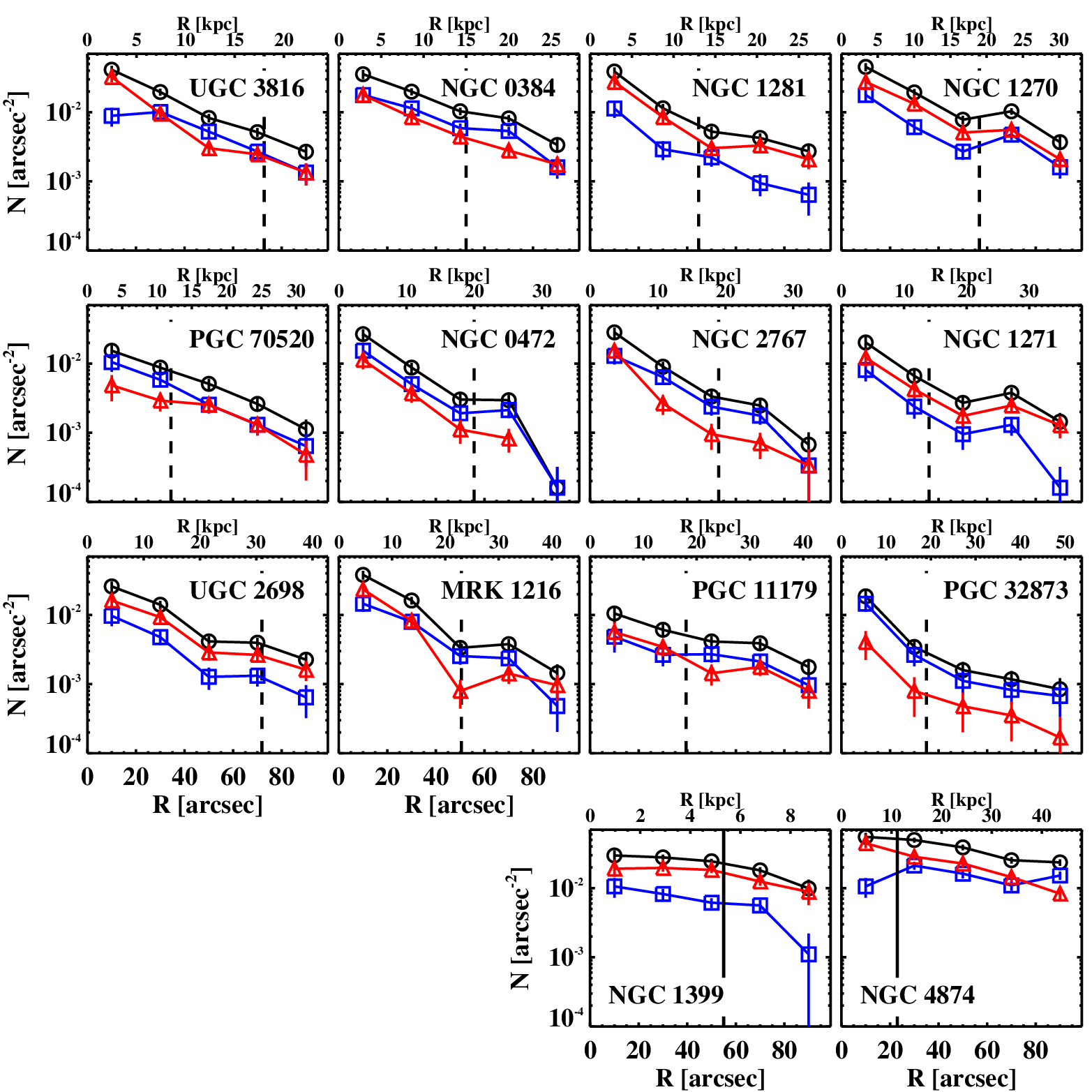}
\caption{Radial number density profiles of the bright sources detected in each target MCEGs and the two reference galaxies.  
Blue rectangles are profiles of blue GCs, 
red triangles are profiles of red GCs, 
and black circles are profiles of all GCs (blue plus red GCs).
The dashed lines denote $R=10R_{e,circ}$ of each galaxy.
For the case of the two reference galaxies, vertical lines denote $R=0.5R_e$.
}
\label{fig_map2}
\end{figure*}

\subsection{Spatial and Radial Distributions of the GCs in the Local MCEGs}

In Figure \ref{fig_map1} 
we show the spatial distribution of the blue and red GCs in each of the target MCEGs. 
The GCs in each of the MCEGs show a central concentration, 
which implies that most of the GCs we select are indeed the members of each galaxy.
We restrict the spatial range for selecting GCs at $R<10R_{e,circ}$, 
similar to the value used for NGC 1277 in \citet{bea18},  
to compare the red GC fraction equivalently.

We also show the spatial distribution of the GCs in the two reference galaxies. Because they are much larger than the MCEGs, only the central region is covered by the HST images so it is hard to see a central concentration clearly. 
In the figure, large solid circles around the two reference galaxies denote $R=0.5R_e$ while large dashed circles around the target MCEGs denote $R=10R_e$. 

In Figure \ref{fig_map2} 
we show the radial number density profiles of all GCs as well as blue and red subpopulations. 
The very central regions of the MCEGs are incomplete so we mask the inner regions at $R < 2\farcs5$ when deriving the radial profile for each MCEG. 
In general, the radial number density of the GCs in the MCEGs decreases as the projected galactocentric distance increases, 
and this indicates that most of the GCs we selected are indeed the members of each galaxy.
All the fields cover $R>10R_{e,circ}$ so the coverage is expected to be wide enough to select most of 
the GCs belonging to the galaxies. 
However, most of the profiles decrease continuously because they are not wide enough to measure the stable background number density.

We also show the radial distribution of the GCs in the two reference galaxies. In general we can see the difference between red GC profile and blue GC profile in giant ETGs, but it is not shown here because these two galaxies are not fully covered by the HST images. 

\section{Discussion} \label{discuss}

\subsection{
Mass-Size Relations of the Local MCEGs in Comparison with the Virgo, Fornax, and Coma ETGs}

\begin{figure*}
\centering
\includegraphics[scale=0.8]{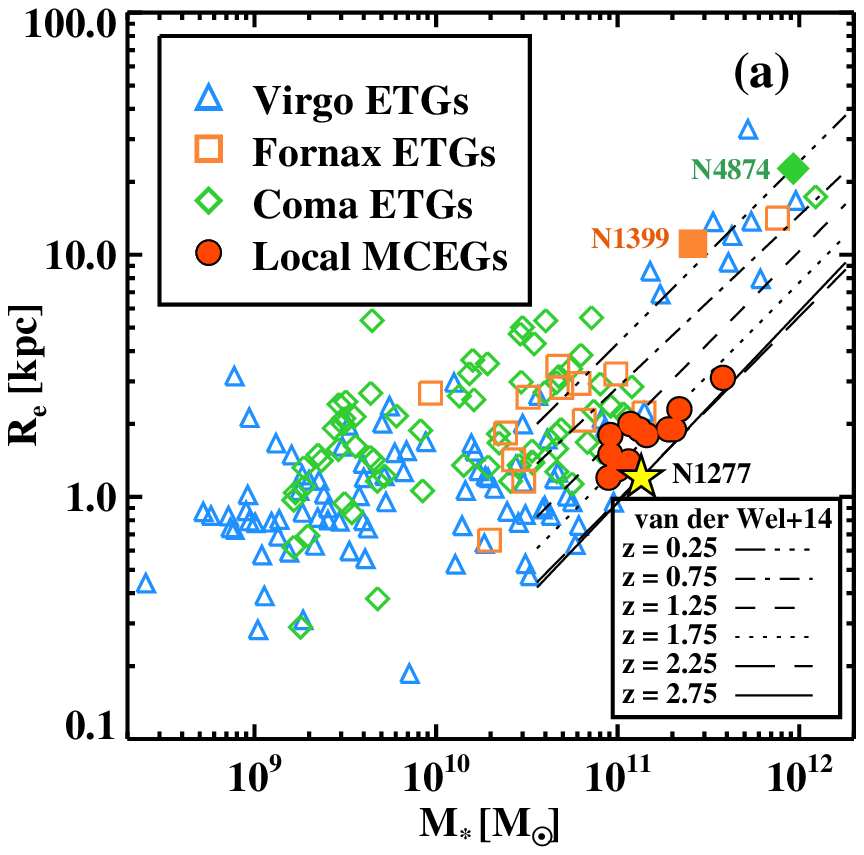}
\includegraphics[scale=0.8]{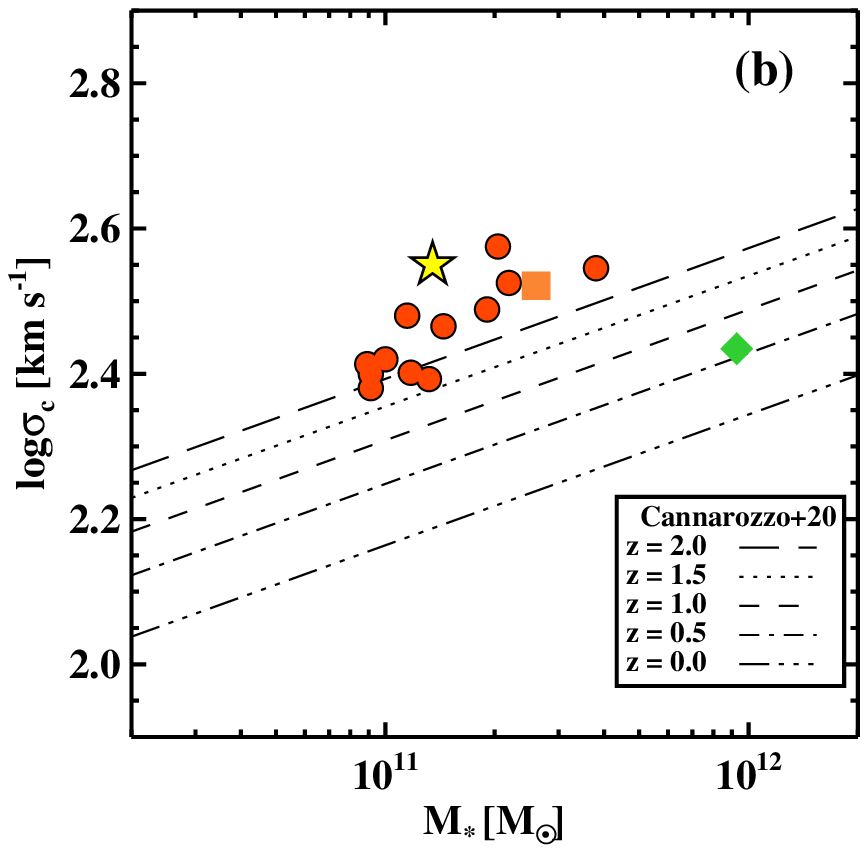}
\caption{
(a) Effective radii ($R_{\rm e}$) vs. stellar masses ($M_*$) 
for the target MCEGs in this study (filled red circles) in comparison with
the Virgo ETGs \citep[blue triangles,][ACSVCS]{fer06,pen08}, 
the Fornax ETGs \citep[orange squares,][ACSFCS]{liu19}, and 
the Coma ETGs \citep[green diamonds,][ACSCCS]{wei14}. 
Filled orange square and 
filled green diamond denote NGC 1399 and NGC 4874. 
The yellow star denotes NGC 1277 \citep{bea18}. 
Black lines are mass-size relations of ETGs at different redshifts \citep{van14}. 
(b) velocity dispersions (log$\sigma_c$) vs. stellar masses ($M_*$) for the target MCEGs, NGC 1277, and the two reference galaxies. Black lines are mass-velocity dispersion relations of ETGs at different redshifts \citep{can20}.
}
\label{fig_ReMs}
\end{figure*}

In Figure \ref{fig_ReMs}(a) 
we display effective radii ($R_{\rm e}$) vs. stellar masses ($M_*$) of our target MCEGs 
in comparison with the Virgo ETGs in the ACS Virgo Cluster Survey \citep[ACSVCS;][]{fer06,pen08}, 
the Fornax ETGs in the ACSFCS \citep{liu19}, 
and the Coma ETGs in the ACSCCS \citep{wei14}. 
We also plot the empirical mass-size relation for $z=0-3$ given by \citet{van14} 
for comparison. 
To compare the stellar mass properly, we consider the mass-to-light ratio (M/L) difference among adopted IMFs: M/L (Salpeter) = 1.6 M/L (Kroupa) = 1.8 M/L (Chabrier). Therefore, we increase the stellar masses of ACSVCS ETGs and mass-size relation by 1.8 where the Chabrier IMF was adopted. We also increase the stellar masses of ACSCCS ETGs by 1.6 where the Kroupa IMF was adopted. 

Distinguishable features in this figure are as follows.
First, all the ETGs in Virgo, Fornax, and Coma follow a well-known 
mass-size relation 
in the sense that the effective radii of the ETGs increase 
as their stellar masses increase. 
In Virgo and Fornax, most galaxies with low stellar mass (i.e. $M_* < 10^{11} M_\sun$) 
have smaller effective radii of $R_e < 3$ kpc, 
while those with high stellar mass (i.e. $M_* > 10^{11} M_\sun$) 
have much larger effective radii of $R_e > 7$ kpc.
The ETGs in Coma follow a similar trend to that of Virgo or Fornax: 
most galaxies with low stellar mass 
have smaller effective radii of $R_e < 5$ kpc, 
while those with high stellar mass 
have much larger effective radii of $R_e > 17$ kpc.

Second, there is an empty region at $M_* > 10^{11} M_\sun$ and $R_e < 7$ kpc 
where almost no Virgo, Fornax, or Coma ETGs are found. 
This shows that there are almost no MCEGs in such clusters. 
The MCEGs in our sample occupy the above empty region. 
NGC 1277 is also included in the same region. 
The MCEGs also follow a similar trend in the 
mass-size relation, 
but with a significant offset in effective radii from those of the massive ETGs: 
they are much smaller than the massive ETGs with similar stellar mass. 
The mass-size relation of the MCEGs is consistent with the relation for $z\approx2$ given by \citet{van14}, as noted in \citet{yil17}. 

In summary, the MCEGs in our sample might have evolved via different routes 
compared with the massive ETGs in Virgo, Fornax, or Coma: 
the former might have grown much less in size than the latter since their formation.
It may be possible that they follow the different mass-size relation
because our sample of MCEGs are located in less dense environments than the cluster ETGs we have compared.
However, about half of our sample of MCEGs are also located in cluster environment: 
NGC 1270, NGC 1271, NGC 1281, 
and UGC 2698 are members of the Perseus cluster (Abell 426) with NGC 1277, 
and PGC 11179 is a member of Abell 400. 
We will further discuss the environment of our sample of MCEGs in the next section.

In Figure \ref{fig_ReMs}(b) we display central velocity dispersion (log$\sigma_c$) vs. stellar mass ($M_*$) of our target MCEGs, NGC 1277, and the two reference galaxies. We also plot the empirical mass-velocity dispersion relation for $z=0-2$ given by \citet{can20} for comparison. 
Here we increase the stellar mass of mass-dispersion relation by 1.8 where the Chabrier IMF was adopted. 
The relation of the MCEGs is consistent with the relation for $z\approx2$. 

Thus, the MCEGs follow the mass-size relation and mass-velocity dispersion relation for galaxies at $z\approx2$. 
This supports that the MCEGs in this study are strong candidates for relic galaxies. 

\subsection{Red GC Fractions vs. Galaxy Properties}

\begin{figure*}
\centering
\includegraphics[scale=0.8]{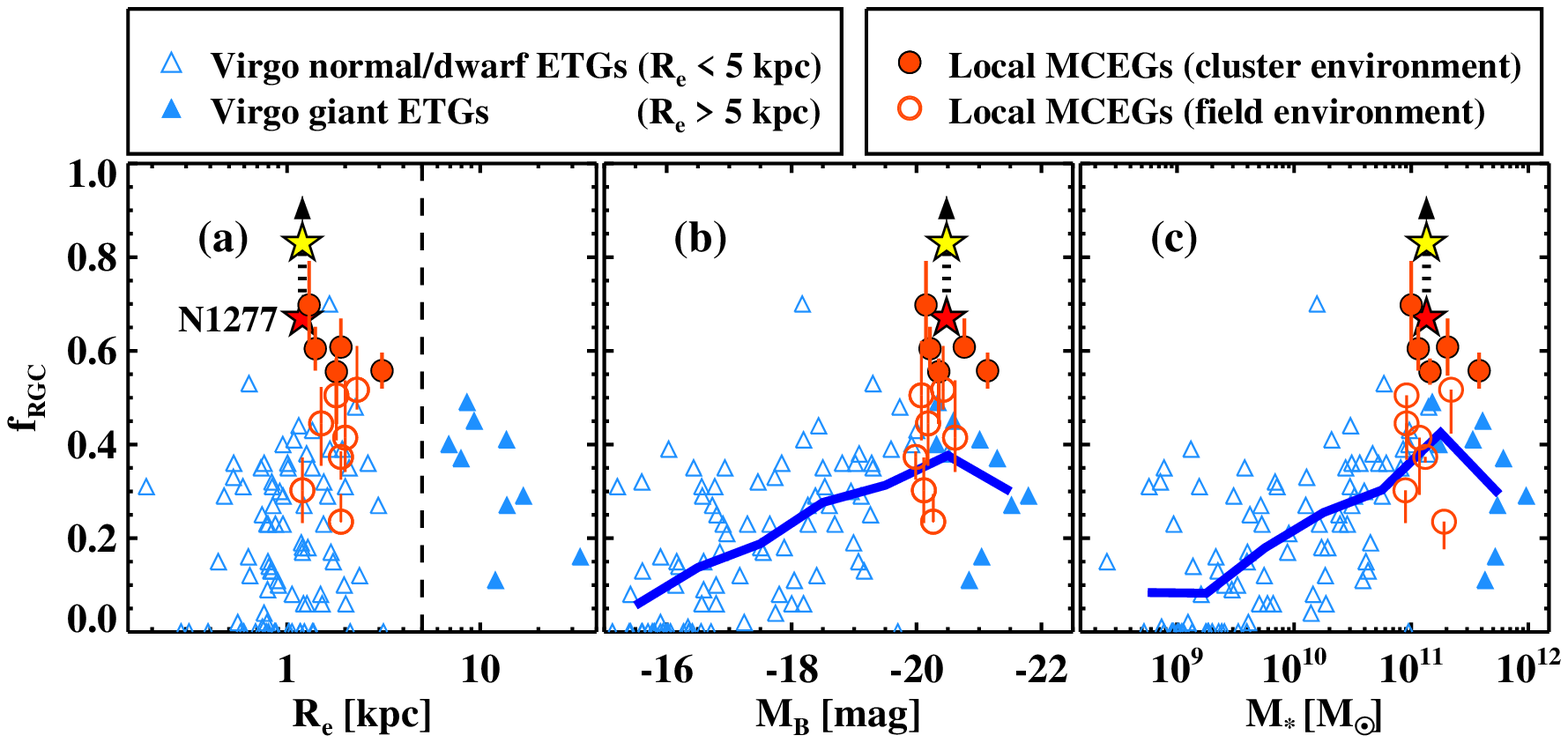}
\caption{
Red GC fractions ($f_{RGC}$) vs. 
(a) effective radii ($R_e$),
(b) B-band absolute magnitudes ($M_B$), and
(c) stellar masses ($M_*$)
for the target MCEGs in this study in comparison with
the Virgo ETGs \citep[][ACSVCS]{cot04,pen06,pen08}.
Symbols are the same as Figure \ref{fig_ReMs} 
but 
we mark the Virgo ETGs with $R_{\rm e}>5$ kpc as filled symbols. 
Filled red circles mark the MCEGs in 
cluster environment and open red circles mark the MCEGs in field environment.
The yellow star represents the value for NGC 1277 given in \citet{bea18}, and the red star denotes the value derived using the fixed color criterion as in this study. 
The blue solid lines in (b) and (c) represent the median of the red GC fractions for the Virgo ETGs 
along with the absolute magnitudes and stellar masses. 
}
\label{fig_frgcMB}
\end{figure*}

For the comparison of red GC fractions in our sample of MCEGs, 
we use the results for the Virgo ETGs in the ACSVCS sample \citep{pen08}. 
They are based on the analysis of the 
HST/ACS data for a large number of galaxies.
For the calculation of the fractions of blue and red GCs in each galaxy of the ACSVCS sample, 
\citet{pen08} applied a hybrid approach to divide the GC sample of a galaxy 
into a blue subpopulation and a red subpopulation. 
For the case of the brightest 21 galaxies (with $M_B<-19.1$ mag) 
they used a dip between the two Gaussians in the KMM two-Gaussian fits 
for the $(g_{475}-z_{850})$ distribution of the GCs. 
They adopted a homoscedastic 
option in the KMM fits. 
For the remaining fainter galaxies they adopted a fixed color of $(g_{475}-z_{850})_0=1.16$. 
Moreover, \citet{pen08} integrated the radial profile of the GCs in each galaxy when counting the number of GCs to prevent overestimating the red GC fraction.
We also use the data for effective radii, B-band absolute magnitudes, and stellar masses 
of the host galaxies in the ACSVCS given by \citet{cot04,fer06,pen08}. 
The total number of host galaxies is 100 for the ACSVCS. 


We did not include NGC 1399 and NGC 4874 
for the comparison with the MCEGs in this study 
because these two galaxies were not fully covered by the HST images used in this study. 
They are large galaxies so only their central regions were covered by the HST images 
used in this study \citep[and previous studies,][]{bla12,cho16}: 
only $R<1\farcm5 \sim1R_e$ region for NGC 1399 and $R<1\farcm6 \sim2R_e$ region for NGC 4874. 
In general, the red GCs in giant ETGs show a stronger central concentration than the blue GCs 
so the red GC fractions derived from the central region in this study (or previous studies) are larger 
than the values derived from the entire regions, as used in the sample MCEGs and Virgo galaxies in this study. 

In Figure \ref{fig_frgcMB}(a) 
we display red GC fractions against 
effective radii 
of the host galaxies in our sample of MCEGs, 
in comparison with the ACSVCS 
ETGs. 
Similarly we show red GC fractions against B-band absolute magnitudes and stellar masses 
of the same galaxies 
in Figures \ref{fig_frgcMB}(b) and (c).
We also plot NGC 1277 using the value of the red GC fraction given by \citet{bea18}.
NGC 1277 has a brighter companion galaxy NGC 1278, 
part of which was covered by the HST/ACS field of NGC 1277.
\citet{bea18} subtracted the contamination of NGC 1278 GCs 
to derive the color distribution of NGC 1277 GCs.
Then they derived a value of the red GC fraction 
from the $(g_{475}-z_{850})$ distribution of NGC 1277 GCs using GMM, 
presenting $f_{RGC}>0.83$ (see their Extended Data Figure 4). 

Several remarkable features are noted in Figure \ref{fig_frgcMB}.
First, we can divide the Virgo ETGs into two groups according to their sizes: 
giant ETGs with $R_{\rm e}>5$ kpc and normal/dwarf ETGs with $R_{\rm e}<5$ kpc.
Our local MCEGs show similar luminosity or mass with those of giant ETGs 
while their sizes are similar to those of normal/dwarf ETGs. 

Second, the red GC fractions of the MCEGs are 
larger than those of the giant Virgo ETGs with similar luminosity or mass. 
The fractions of the MCEGs are also 
larger than those of the normal/dwarf Virgo ETGs with similar size.
We also mark 
the median of the red GC fractions for the Virgo ETGs
along the magnitude and stellar mass, 
with a magnitude bin of $\Delta M_B=1.0$ mag and stellar mass bin of $\Delta \rm{log}(M_*/M_\odot)=0.5$.
Following the overall trend, 
we can see that the red GC fractions of the Virgo ETGs increase as their luminosity or mass increase.
However, they begin to decrease at $M_B\approx -20.5$ mag or 
$M_*\approx 2\times10^{11}M_\sun$ 
as mentioned in \citet{pen08}.
In contrast, the red GC fractions of our target MCEGs 
do not decrease even after $M_B\approx -20.5$ mag or 
$M_*\approx 2\times10^{11}M_\sun$. 

Third, we can divide the MCEG sample into two subsamples: one for the cluster MCEGs and the other for the group/field MCEGs (field MCEGs hereafter). The cluster MCEG subsample includes five MCEGs, and the field subsample has seven. Note that four out of the five MCEGs in clusters are located very close to the center ($R<250$ kpc) of their host clusters, so they are located in a denser environment compared with most galaxies in Virgo. 
The mean red GC fraction of our 12 MCEG targets is $f_{RGC}=0.48\pm0.14$. 
Interestingly, if we restrict our sample to 
cluster MCEGs, 
this value increases to 
$f_{RGC}=0.60\pm0.06$. 
This value is
$\sim$0.3 larger than the mean red GC fraction of the giant Virgo ETGs, $f_{RGC}=0.33\pm0.13$.
If we restrict our sample to 
field MCEGs, 
then the mean fraction decreases to 
$f_{RGC}=0.40\pm0.10$. 
This value is 0.2 lower than the value for the cluster MCEGs, and is similar to the value for the Virgo ETGs. 
The red GC fraction for normal-size ETGs in the field may be lower than that for cluster ETGs. However, there are no data available for the field normal-size ETGs as homogeneous as those for Virgo galaxies so we could not include them for comparison in this study. 
These results on red GC fractions are 
consistent with the previous finding that relic galaxies prefer dense environments. The environmental preference of MCEGs is well summarized in \citet{bea18}. In short, local MCEGs, 
like other normal massive ETGs, 
prefer dense environments both in observational and theoretical view \citep[e.g.][]{pog13,dam15,str15,per16}.
If we restrict the area of GC selection for each galaxy to $R<5R_{e,circ}$, the mean red GC fraction does not change much 
($f_{RGC}=0.51\pm0.14$ for all 12 MCEGs,  
$f_{RGC}=0.60\pm0.10$ for the cluster MCEGs and $0.45\pm0.13$ for the field MCEGs). 

Fourth, 
the red GC fractions of the MCEGs are not 
as 
large as 
those of NGC 1277. 
\citet{bea18} obtained $f_{RGC}>0.83$ by applying GMM analysis to the $(g_{475}-z_{850})_0$ color distribution 
of the NGC 1277 GCs. 
If we take the result as it is, 
NGC 1277 can be considered as the unique sample hosting 
the highest fraction of red GCs. 
However, there is a possibility that the red GC fractions of our target MCEGs are rather underestimated 
compared with NGC 1277. 
In the case of NGC 1277, blue and red GC subpopulations are divided at $(g_{475}-z_{850})_0\approx1.0$ 
according to the GMM analysis
in \citet{bea18}. 
This color is much 
bluer than the typical division color, $(g_{475}-z_{850})_0=1.16$, 
which was used for the previous studies of the GC color distributions 
\citep[e.g. ACSVCS ETGs, NGC 1399, and NGC 1278,][]{pen06,bla12,bea18}. 
Therefore, if we use a fixed color of $(g_{475}-z_{850})_0=1.16$ 
as in the previous studies to derive the red GC fraction of NGC 1277 
instead of using GMM analysis, 
we obtain a much smaller fraction of $f_{RGC}=0.67$ 
\citep[see Figure 2 or Extended Data Figure 4 in][]{bea18}. This value is similar to the value for NGC 1281 we derived in this study, $f_{RGC}=0.70\pm0.09$.
Thus, it is possible that the red GC fraction derived from a fixed $(I_{814}-H_{160})$ color 
for our sample may be an underestimate compared with NGC 1277. 
If there is a galaxy such as NGC 1277, 
hosting a 
large fraction of red GCs with a blue tail in their color distribution, 
using a fixed color can underestimate its red GC fraction. 
This implies that if we apply GMM analysis to our sample of MCEGs, we may obtain 
larger red GC fractions similar to NGC 1277. 
In the case of NGC 1277, there is about 0.2 difference (0.67 and 0.83) in the red GC fraction 
between the two methods (color cut and GMM). 
Therefore, the red GC fraction of MCEGs may systematically increase by up to $\sim$0.2. 
Then NGC 1277 
may not be a unique sample anymore.

\subsection{Origins of GCs and Relic Galaxies}



In this study we find that nearby MCEGs have GC systems dominated by red GCs, 
and their red GC fractions are on average 
larger than those of 
giant ETGs with similar stellar mass or normal/dwarf ETGs with similar size. 
This result holds significant implications on the origins of GCs and relic galaxies.

\subsubsection{Origin of Blue and Red GCs}

The fraction of metal-rich GCs in local massive galaxies increases 
from $\sim$20\% to $\sim$80\% 
as the galactocentric distance decreases 
from the outer region to the central region \citep[see Fig. 19 in][]{har17},
while the GCs found in the outer halos of local massive ETGs or intracluster regions are mostly metal-poor 
\citep{pen06,lee10a,pen11,har17,for18}. 
These results lead to the following scenario on the origin of GCs:
Metal-rich GCs are formed either in the central region of massive galaxies 
during dissipative collapse of their progenitor or in the major merger of massive progenitors. 
On the other hand, most metal-poor GCs are formed in dwarf galaxies 
\citep[e.g.][and references therein]{cot98,bro06,lee10b,for18}.

In this study we find that most 
GCs in MCEGs are red GCs (i.e. metal-rich GCs).
These red GCs must have originated from the progenitors of the MCEGs 
rather than from being transferred from other satellite galaxies.
Thus, the presence of dominant red GC populations in MCEGs is 
clear evidence 
which shows that 
red GCs are formed mainly in massive galaxies. 
Very low fraction of blue GCs in MCEGs 
and very high fractions of blue GCs in the outer region of local massive ETGs 
imply that the major origin of blue GCs is low mass dwarf galaxies 
which were accreted to the massive galaxies.

\subsubsection{Are Local MCEGs 
Relic Galaxies?}

It is expected that most red nugget galaxies in high density environment (i.e. galaxy clusters or groups) 
experience many mergers of diverse types during their evolution, 
and they grow not only in their size and stellar mass but also in 
their blue GC fraction.
In this study we find not only that red GC fractions of the MCEGs in our sample 
are higher than those of local massive ETGs with similar stellar mass, 
but also that they are higher than those of local normal/dwarf ETGs with similar size.
This implies that the progenitors of the MCEGs became red nugget galaxies 
of which GC systems are dominated by red GCs about 12 Gyr ago 
\citep[assuming that they are as old as the metal-rich GCs in the MW,][]{oli20}, 
and they must have undergone very few mergers later. 
They grew little not only in size and stellar mass but also in 
blue GC fraction.  
Thus, we conclude that the MCEGs in our sample are 
genuine relic galaxies.

\section{Summary and Conclusion}

We present a search for GCs in 
12 nearby MCEGs, 
taking advantage of the high resolution HST/WFC3 F814W/F160W images in the archive. 
The 
mass-size relation of these MCEGs shows a significant difference 
from those of local massive ETGs in Virgo, Fornax, and Coma, as shown in Figure \ref{fig_frgcMB}, 
but it is consistent with that of red nugget galaxies at high redshift ($z\approx 2$) \citep{van14,yil17}. 
To see whether these MCEGs host GC systems different from those of local massive ETGs as well, 
we investigate various photometric properties of the GC systems in these MCEGs.
Primary results and conclusion are summarized as follows.

\begin{enumerate}
\item CMDs show that most MCEGs host a rich population of GCs. Absolute magnitudes of the detected GCs range 
from $M_{I} \approx -13.0$  to --8.5 mag, 
the faint limit of which varies on the distance to their host galaxies.

\item Color distributions of the GC candidates with $I_{814,0}<25.0$ mag show that most MCEGs show a dominant broad component at $0.0<(I_{814}-H_{160})_0<1.1$, which is mainly composed of GCs in each galaxy.

\item We estimate the fraction of red GCs using a fixed color criterion of $(I_{814}-H_{160})_0 = 0.49\pm0.04$. 
Red GC fractions of the MCEGs 
are about 0.2 higher than those of the giant Virgo ETGs with similar stellar mass. 
Some 
MCEGs are expected to show comparably 
large red GC fractions like NGC 1277, 
a known relic galaxy \citep{bea18}.


\item These results imply that a majority of red GCs were formed early in massive galaxies 
and that most MCEGs have undergone very few mergers after they became red nuggets about 10 Gyr ago. 
Thus, it is concluded that they are genuine relic galaxies.

\end{enumerate}



It is expected that more MCEGs will be discovered from new surveys. 
Recently, \citet{tor18}  and \citet{sco20} found 37 new Ultra Compact Massive Galaxies (UCMGs) 
($M_* >8 \times 10^{10} M_\sun$ and $R_e < 1.5$ kpc) at $z<0.5$ from the Kilo Degree Survey. 
They combined them with 55 UCMGs 
from the literature, 
producing the largest spectroscopically confirmed sample of UCMGs at $z<0.5$. 
Note that a rich population of GCs was found 
even in distant galaxy clusters like Abell 2744 at redshift of $z=0.308$ 
from deep HST F814W/F105W images \citep{lee16b}.
The UCMGs and new MCEGs at $z<0.5$ will be good targets for future studies of their GC systems 
with HST or JWST.
Moreover, follow-up spectroscopic studies for the GCs in the MCEGs with a large telescope (e.g. Keck, GMT, TMT, or eELT) 
will be helpful to confirm the membership of the GCs and to study their kinematic properties. 

\acknowledgements
We thank the anonymous referee for useful comments. 
J.K. was supported by the Global Ph.D. Fellowship Program (NRF-2016H1A2A1907015) of the National Research Foundation (NRF).
This work was supported by the NRF funded by the Korean Government (MSIT) (NRF-2019R1A2C2084019).
We thank Brian S. Cho for improving the English in the manuscript.


\clearpage






\begin{thebibliography}{}


\bibitem[Alamo-Mart{\'\i}nez et al.(2021)]{ala21} Alamo-Mart{\'\i}nez, K.~A., Chies-Santos, A.~L., Beasley, M.~A., et al.\ 2021, \mnras 

\bibitem[Almaini et al.(2017)]{alm17} Almaini, O., Wild, V., Maltby, D.~T., et al.\ 2017, \mnras, 472, 1401 


\bibitem[Ashman et al.(1994)]{ash94} Ashman, K.~M., Bird, C.~M., \& Zepf, S.~E.\ 1994, \aj, 108, 2348 




\bibitem[Beasley et al.(2018)]{bea18} Beasley, M.~A., Trujillo, I., Leaman, R., \& Montes, M.\ 2018, \nat, 555, 483 

\bibitem[Bertin \& Arnouts(1996)]{ber96} Bertin, E., \& Arnouts, S.\ 1996, \aaps, 117, 393

\bibitem[Blakeslee et al.(2009)]{bla09} Blakeslee, J.~P., Jord{\'a}n, A., Mei, S., et al.\ 2009, \apj, 694, 556

\bibitem[Blakeslee et al.(2012)]{bla12} Blakeslee, J.~P., Cho, H., Peng, E.~W., et al.\ 2012, \apj, 746, 88 NGC 1399

\bibitem[Bressan et al.(2012)]{bre12} Bressan, A., Marigo, P., Girardi, L., et al.\ 2012, \mnras, 427, 127

\bibitem[Brodie \& Strader(2006)]{bro06} Brodie, J.~P., \& Strader, J.\ 2006, \araa, 44, 193 



\bibitem[Cannarozzo et al.(2020)]{can20} Cannarozzo, C., Sonnenfeld, A., \& Nipoti, C.\ 2020, \mnras, 498, 1101

\bibitem[Carter et al.(2008)]{car08} Carter, D., Goudfrooij, P., Mobasher, B., et al.\ 2008, \apjs, 176, 424 

\bibitem[Cho et al.(2016)]{cho16} Cho, H., Blakeslee, J.~P., Chies-Santos, A.~L., et al.\ 2016, \apj, 822, 95 

\bibitem[C\^ot\'e et al.(1998)]{cot98} C\^ot\'e, P., Marzke, R.O., \& West, M.J. 1998, \apj, 501, 554

\bibitem[C{\^o}t{\'e} et al.(2004)]{cot04} C{\^o}t{\'e}, P., Blakeslee, J.~P., Ferrarese, L., et al.\ 2004, \apjs, 153, 223

\bibitem[Damjanov et al.(2015)]{dam15} Damjanov, I., Zahid, H.~J., Geller, M.~J., et al.\ 2015, \apj, 815, 104

\bibitem[Dekel \& Burkert(2014)]{dek14} Dekel, A., \& Burkert, A.\ 2014, \mnras, 438, 1870 



\bibitem[Ferrarese et al.(2006)]{fer06} Ferrarese, L., C{\^o}t{\'e}, P., Jord{\'a}n, A., et al.\ 2006, \apjs, 164, 334

\bibitem[Ferr{\'e}-Mateu et al.(2017)]{fer17} Ferr{\'e}-Mateu, A., Trujillo, I., Mart{\'{\i}}n-Navarro, I., et al.\ 2017, \mnras, 467, 1929  


\bibitem[Forbes \& Remus(2018)]{for18} Forbes, D.~A., \& Remus, R.-S.\ 2018, \mnras, 479, 4760 

\bibitem[Furlong et al.(2017)]{fur17} Furlong, M., Bower, R.~G., Crain, R.~A., et al.\ 2017, \mnras, 465, 722 

\bibitem[Girardi et al.(2005)]{gir05} Girardi, L., Groenewegen, M.~A.~T., Hatziminaoglou, E., et al.\ 2005, \aap, 436, 895


\bibitem[Gonzaga et al.(2012)]{gon12} Gonzaga, S., Hack, W., Fruchter, A., Mack, J., eds. 2012, The DrizzlePac Handbook. (Baltimore, STScI)

\bibitem[Harris(2009)]{har09} Harris, W.~E.\ 2009, \apj, 699, 254 


\bibitem[Harris et al.(2017)]{har17} Harris, W.~E., Ciccone, S.~M., Eadie, G.~M., et al.\ 2017, \apj, 835, 101 


\bibitem[Illingworth et al.(2013)]{ill13} Illingworth, G.~D., Magee, D., Oesch, P.~A., et al.\ 2013, \apjs, 209, 6




\bibitem[Jord{\'a}n et al.(2007)]{jor07} Jord{\'a}n, A., Blakeslee, J.~P., C{\^o}t{\'e}, P., et al.\ 2007, \apjs, 169, 213


\bibitem[Khochfar \& Silk(2006)]{kho06} Khochfar, S., \& Silk, J.\ 2006, \apjl, 648, L21 

\bibitem[Lapi et al.(2018)]{lap18} Lapi, A., Pantoni, L., Zanisi, L., et al.\ 2018, \apj, 857, 22 

\bibitem[Lee et al.(2010a)]{lee10a} Lee, M.~G., Park, H.~S., \& Hwang, H.~S.\ 2010, Science, 328, 334 

\bibitem[Lee et al.(2010b)]{lee10b} Lee, M.~G., Park, H.~S., Hwang, H.~S., et al.\ 2010, \apj, 709, 1083 

\bibitem[Lee \& Jang(2016a)]{lee16a} Lee, M.~G., \& Jang, I.~S.\ 2016, \apj, 819, 77

\bibitem[Lee \& Jang(2016b)]{lee16b} Lee, M.~G., \& Jang, I.~S.\ 2016, \apj, 831, 108 

\bibitem[Lee et al.(2018)]{lee18} Lee, M.~G., Kang, J., \& Im, M.\ 2018, \apjl, 859, L6

\bibitem[Lee, Jang, \& Kang (2019)]{lee19} Lee, M.~G., Jang, I.~S., \& Kang, J. \ 2019, \apj, in press 



\bibitem[Liu et al.(2019)]{liu19} Liu, Y., Peng, E.~W., Jord{\'a}n, A., et al.\ 2019, \apj, 875, 156




\bibitem[Muratov \& Gnedin(2010)]{mur10} Muratov, A.~L., \& Gnedin, O.~Y.\ 2010, \apj, 718, 1266 

\bibitem[Naab \& Ostriker(2017)]{nab17} Naab, T., \& Ostriker, J.~P.\ 2017, \araa, 55, 59 



\bibitem[Oliveira et al.(2020)]{oli20} Oliveira, R.~A.~P., Souza, S.~O., Kerber, L.~O., et al.\ 2020, \apj, 891, 37

\bibitem[Oser et al.(2010)]{ose10} Oser, L., Ostriker, J.~P., Naab, T., Johansson, P.~H., \& Burkert, A.\ 2010, \apj, 725, 2312 

\bibitem[Oser et al.(2012)]{ose12} Oser, L., Naab, T., Ostriker, J.~P., \& Johansson, P.~H.\ 2012, \apj, 744, 63 


\bibitem[Peng et al.(2006)]{pen06} Peng, E.~W., Jord{\'a}n, A., C{\^o}t{\'e}, P., et al.\ 2006, \apj, 639, 95 

\bibitem[Peng et al.(2008)]{pen08} Peng, E.~W., Jord{\'a}n, A., C{\^o}t{\'e}, P., et al.\ 2008, \apj, 681, 197

\bibitem[Peng et al.(2011)]{pen11} Peng, E.~W., Ferguson, H.~C., Goudfrooij, P., et al.\ 2011, \apj, 730, 23 

\bibitem[Peralta de Arriba et al.(2016)]{per16} Peralta de Arriba, L., Quilis, V., Trujillo, I., et al.\ 2016, \mnras, 461, 156


\bibitem[Poggianti et al.(2013)]{pog13} Poggianti, B.~M., Calvi, R., Bindoni, D., et al.\ 2013, \apj, 762, 77


\bibitem[Schlafly \& Finkbeiner(2011)]{sch11} Schlafly, E.~F., \& Finkbeiner, D.~P.\ 2011, \apj, 737, 103

\bibitem[Scognamiglio et al.(2020)]{sco20} Scognamiglio, D., Tortora, C., Spavone, M., et al.\ 2020, \apj, 893, 4 

\bibitem[Secker(1995)]{sec95} Secker, J.\ 1995, \pasp, 107, 496



\bibitem[Spiniello et al.(2020)]{spi20} Spiniello, C., Tortora, C., D'Ago, G., et al.\ 2020, arXiv:2011.05347

\bibitem[Stringer et al.(2015)]{str15} Stringer, M., Trujillo, I., Dalla Vecchia, C., et al.\ 2015, \mnras, 449, 2396


\bibitem[Toft et al.(2014)]{tof14} Toft, S., Smol{\v c}i{\'c}, V., Magnelli, B., et al.\ 2014, \apj, 782, 68 

\bibitem[Toft et al.(2017)]{tof17} Toft, S., Zabl, J., Richard, J., et al.\ 2017, \nat, 546, 510 

\bibitem[Tortora et al.(2018)]{tor18} Tortora, C., Napolitano, N.~R., Spavone, M., et al.\ 2018, \mnras, 481, 4728 


\bibitem[Trujillo et al.(2014)]{tru14} Trujillo, I., Ferr{\'e}-Mateu, A., Balcells, M., Vazdekis, A., \& S{\'a}nchez-Bl{\'a}zquez, P.\ 2014, \apjl, 780, L20 

\bibitem[van den Bosch et al.(2015)]{van15} van den Bosch, R.~C.~E., Gebhardt, K., G{\"u}ltekin, K., Y{\i}ld{\i}r{\i}m, A., \& Walsh, J.~L.\ 2015, \apjs, 218, 10 

\bibitem[van der Wel et al.(2014)]{van14} van der Wel, A., Franx, M., van Dokkum, P.~G., et al.\ 2014, \apj, 788, 28 

\bibitem[van Dokkum et al.(2010)]{van10} van Dokkum, P.~G., Whitaker, K.~E., Brammer, G., et al.\ 2010, \apj, 709, 1018 





\bibitem[Weinzirl et al.(2014)]{wei14} Weinzirl, T., Jogee, S., Neistein, E., et al.\ 2014, \mnras, 441, 3083

\bibitem[Wellons et al.(2016)]{wel16} Wellons, S., Torrey, P., Ma, C.-P., et al.\ 2016, \mnras, 456, 1030 




\bibitem[Y{\i}ld{\i}r{\i}m et al.(2015)]{yil15} Y{\i}ld{\i}r{\i}m, A., van den Bosch, R.~C.~E., van de Ven, G., et al.\ 2015, \mnras, 452, 1792 

\bibitem[Y{\i}ld{\i}r{\i}m et al.(2016)]{yil16} Y{\i}ld{\i}r{\i}m, A., van den Bosch, R.~C.~E., van de Ven, G., et al.\ 2016, \mnras, 456, 538 

\bibitem[Y{\i}ld{\i}r{\i}m et al.(2017)]{yil17} Y{\i}ld{\i}r{\i}m, A., van den Bosch, R.~C.~E., van de Ven, G., et al.\ 2017, \mnras, 468, 4216 



\end{thebibliography}
\end{document}